# Next-Generation Quantum Theory of Atoms in Molecules for the Ground and Excited States of Fulvene

Wei Jie Huang[1], Roya Momen[1], Alireza Azizi[1], Tianlv Xu[1], Steven R. Kirk[1*], Michael Filatov[1,2] and Samantha Jenkins[1*]

[1]*Key Laboratory of Chemical Biology and Traditional Chinese Medicine Research and Key Laboratory of Resource Fine-Processing and Advanced Materials of Hunan Province of MOE, College of Chemistry and Chemical Engineering, Hunan Normal University, Changsha, Hunan 410081, China*

[2]*Department of Chemistry, Ulsan National Institute of Science and Technology (UNIST), 50 UNIST-gil, Ulsan 44919, Korea*

email: steven.kirk@cantab.net
email: mike.filatov@gmail.com
email: samanthajsuman@gmail.com

A vector-based representation of the chemical bond is introduced, which we refer to as the bond-path frame-work set $\mathbb{B} = \{p, q, r\}$, where $p$, $q$ and $r$ represent three eigenvector-following paths with corresponding lengths $\mathbb{H}^*$, $\mathbb{H}$ and the familiar quantum theory of atoms in molecules (QTAIM) bond-path length. The intended application of $\mathbb{B}$ is for molecules subjected to various types of reactions and distortions, including photo-isomerization reactions, applied torsions θ, or normal modes of vibration. The lengths $\mathbb{H}^*$ and $\mathbb{H}$ of the eigenvector-following paths are constructed using the $\underline{\mathbf{e}}_1$ and $\underline{\mathbf{e}}_2$ Hessian eigenvectors respectively along the bond-path, these corresponding to the least and most preferred directions of charge density accumulation. In particular, the paths $p$ and $q$ provide a vector representation of the scalar QTAIM ellipticity ε. The bond-path frame-work set $\mathbb{B}$ is applied to the excited state deactivation of fulvene that involves distortions along various intramolecular degrees of freedom, such as the bond stretching/compression of bond length alternation (BLA) and bond torsion distortions. We find that the $\mathbb{H}^*$ and $\mathbb{H}$ lengths can differentiate between the ground and excited electronic states, in contrast to the QTAIM bond-path length. Five unique paths were presented for $\mathbb{B} = \{(p_0,p_1), (q_0,q_1), r\}$ for the ground and first excited states where the profile of the scaling factor, the ellipticity ε, reveals a large unexpected asymmetry for the excited state.

# 1. Introduction

In this work we propose 'next-generation' quantum theory of atoms in molecules (QTAIM)[1] to analyze all types of bonding, including single and multiple bonds in molecules subjected to various types of reactions and distortions, including photo-isomerization reactions, applied torsions θ, or normal modes of vibration. The form of this next-generation development is the bond-path framework set $\mathbb{B}$ which is a vector-based representation of the chemical bond within the QTAIM framework.

Existing notable representations of the chemical bond, also within the QTAIM framework include the work of Jones and Eberhart and also that of Morgenstern *et al* on bond-bundles in open systems, whereby molecules are partitioned through an extension of QTAIM where bounded regions of space containing non-bonding or lone-pair electrons are created that lead to bond orders consistent with expectation from theories of directed valence[2,3]. Additional insights into the chemical bond were also provided by Popelier[4]. Previously, two of the current authors found that the charge density does not always rotate in accordance with the nuclei for the double bond isomerization of fulvene[5]. Instead, the resistance of the charge density to rotation was found to depend on the electronic state and the behavior of the charge density was locally different for different directions of the bond-path framework.

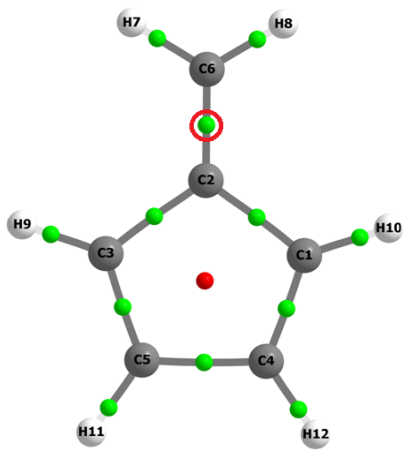

**Scheme 1.** The atom labels correspond to the molecular graph atom numbering scheme and the torsion C2-C6 *BCP* is indicated by a red circle.

Here, we shall investigate the double bond isomerization of fulvene in the ground $S_0$ and first excited $S_1$ electronic states with the goal of finding a QTAIM representation of the chemical bond that can differentiate between ground and excited electronic states, see **Scheme 1**. As we shall see in this investigation the *bond-path*, the current and widely used QTAIM interpretation of the chemical bond, is unable to differentiate between the $S_0$ and $S_1$ electronic states. Starting point for this exploration will be the ellipticity ε, constructed from the $\lambda_1$ and $\lambda_2$ eigenvalues of the Hessian of $\rho(\mathbf{r})$ associated with the accumulation of the charge distribution in chemical bond. The ellipticity ε, where $\varepsilon = |\lambda_1|/|\lambda_2| - 1$, has already been found to distinguish between the $S_0$ and $S_1$ electronic states[6]. This next-generation QTAIM interpretation of the

chemical bond will be constructed using the ellipticity ε and the associated **e₁** and **e₂** eigenvectors for all points along the bond-path.

A detailed exploration of the potential energy surfaces (PESs) will be undertaken along the minimum energy pathways (MEPs) connecting the Franck-Condon (FC) point with planar and torsional conical intersections (CIs) occurring along the torsional coordinate (θ) as well as the bond-length-alternation (BLA) distortion. The BLA is calculated as the difference between the average length of the formally single bonds and the average length of the formally double bonds in the molecule.

## 2. Theory and Methods

*2.1 The QTAIM and stress tensor BCP properties; ellipticity ε, total local energy density $H(r_b)$, stress tensor eigenvalue $\lambda_{3\sigma}$ and stress tensor stiffness, $\mathbb{S}_\sigma$*

We use QTAIM and the stress tensor analysis that utilizes higher derivatives of $\rho(\mathbf{r_b})$ in effect, acting as a 'magnifying lens' on the $\rho(\mathbf{r_b})$ derived properties of the wave-function. We will use QTAIM[1] to identify critical points in the total electronic charge density distribution $\rho(\mathbf{r})$ by analyzing the gradient vector field $\nabla\rho(\mathbf{r})$. These critical points can further be divided into four types of topologically stable critical points according to the set of ordered eigenvalues $\lambda_1 < \lambda_2 < \lambda_3$, with corresponding eigenvectors **e₁**, **e₂**, **e₃** of the Hessian matrix. The Hessian of the total electronic charge density $\rho(\mathbf{r})$ is defined as the matrix of partial second derivatives with respect to the spatial coordinates. These critical points are labeled using the notation (R, ω) where R is the rank of the Hessian matrix, the number of distinct non-zero eigenvalues and ω is the signature (the algebraic sum of the signs of the eigenvalues); the (3, -3) [nuclear critical point (*NCP*), a local maximum generally corresponding to a nuclear location], (3, -1) and (3, 1) [saddle points, called bond critical points (*BCP*) and ring critical points (*RCP*), respectively] and (3, 3) [the cage critical points (*CCP*)]. In the limit that the forces on the nuclei become vanishingly small, an atomic interaction line[7] becomes a bond-path, although not necessarily a chemical bond[8]. The complete set of critical points together with the bond-paths of a molecule or cluster is referred to as the molecular graph.

The eigenvector **e₃** indicates the direction of the bond-path at the *BCP*. The most and least preferred directions of electron accumulation are **e₂** and **e₁**, respectively[9–11]. The ellipticity, ε provides the relative accumulation of $\rho(\mathbf{r_b})$ in the two directions perpendicular to the bond-path at a *BCP*, defined as $\varepsilon = |\lambda_1|/|\lambda_2| - 1$ where $\lambda_1$ and $\lambda_2$ are negative eigenvalues of the corresponding eigenvectors **e₁** and **e₂** respectively. Recently, for the 11-cis retinal subjected to a torsion ±θ, we have recently demonstrated that the **e₂** eigenvector of the torsional *BCP* corresponded to the preferred +θ direction of rotation as defined by the PES profile[12].

Previously, we defined a bond-path stiffness, $\mathbb{S} = \lambda_2/\lambda_3$, as a measure of rigidity of the bond-path[13]. Diagonalization of the stress tensor, σ(**r**), returns the principal electronic stresses. In this work we use the definition of the stress tensor proposed by Bader[14] to investigate the stress tensor properties within the

QTAIM partitioning scheme. Thus, analogously to the QTAIM descriptor, we use the stress tensor stiffness, $\mathbb{S}_\sigma = |\lambda_{1\sigma}|/|\lambda_{3\sigma}|$, that previously was found to be an effective descriptor of the 'resistance' of the bond-path to an applied torsion[5]. Previously, it was found that the stress tensor stiffness, $\mathbb{S}_\sigma$ produced results that were in line with physical intuition[5,13]. The stress tensor eigenvalue $\lambda_{3\sigma}$ is associated with the bond-path and previously values of $\lambda_{3\sigma} < 0$ were found to be associated with transition-type behavior in biphenyl[13] and molecular motors[6] and indicated the bond critical point was close to rupturing and therefore we described as being *unstable*.

The total local energy density $H(\mathbf{r}_b)$[5,15]

$$H(\mathbf{r}_b) = G(\mathbf{r}_b) + V(\mathbf{r}_b), \qquad (1)$$

where $G(\mathbf{r}_b)$ and $V(\mathbf{r}_b)$ are the local kinetic and potential energy densities at a *BCP*, defines a degree of covalent character: A negative $H(\mathbf{r}_b) < 0$ for the closed-shell interaction, $\nabla^2\rho(\mathbf{r}_b) > 0$, indicates a *BCP* with a degree of covalent character and conversely a positive $H(\mathbf{r}_b) > 0$ reveals a lack of covalent character for the closed-shell *BCP*.

*2.2 The QTAIM bond-path properties; BPL, the eigenvector-following paths with lengths $\mathbb{H}$, $\mathbb{H}^*$ and the bond-path framework set $\mathbb{B}$*

The bond-path length (BPL) is defined as the length of the path traced out by the $\underline{\mathbf{e}_3}$ eigenvector of the Hessian of the total charge density $\rho(\mathbf{r})$, passing through the *BCP*, along which $\rho(\mathbf{r})$ is locally maximal with respect to any neighboring paths. The bond-path curvature separating two bonded nuclei is defined as the dimensionless ratio:

$$(BPL - GBL)/GBL, \qquad (2)$$

where BPL is as the associated bond-path length and the geometric bond length GBL is the inter-nuclear separation. The BPL often exceeds the GBL particularly for weak or strained bonds and unusual bonding environments[16]. Earlier, one of the current authors hypothesized that the morphology of a bond-path may be 1-D i.e. a linear bond-path equal in length to the bonded inter-nuclear separation, bent with one radius of curvature (2-D) only in the direction of $\underline{\mathbf{e}_2}$. For 3-D bond-paths, there are minor and major radii of curvature specified by the directions of $\underline{\mathbf{e}_2}$ and $\underline{\mathbf{e}_1}$ respectively.[17] In this investigation we suggest the involvement of the $\underline{\mathbf{e}_3}$ eigenvector also, in the form of a bond-path twist. Bond-paths possessing zero and non-zero values of the bond-path curvature defined by equation **(2)** can be considered to possess 1-D and 2-D topologies respectively. For the realization of this hypothesis we start by choosing the length traced out in 3-D by the path swept by the tips of the scaled $\underline{e}_2$ eigenvectors of the $\lambda_2$ eigenvalue. It was observed during calculations of the $\underline{e}_1$ and $\underline{e}_2$ eigenvectors at successive points along the bond-path that in some cases, these eigenvectors, both being perpendicular to the bond-path tracing eigenvector $\underline{e}_3$, 'switched places'. Any calculation of a vector tip path following, say, the unscaled $\underline{e}_1$ eigenvector would then show a large 'jump' as it swapped directions with the corresponding $\underline{e}_2$ eigenvector. This phenomenon indicates the presence of a location where the corresponding $\lambda_1$ and $\lambda_2$ eigenvalues are degenerate and hence where the ellipticity $\varepsilon$ must be zero. The choice of the ellipticity $\varepsilon$ as scaling factor is motivated by the fact that the scaled vector tip paths drop smoothly onto the bond-path, ensuring that the tip paths are always continuous. Alternative scaling factors were considered that also can take a zero value, included $|\lambda_1-\lambda_2|$ this was not pursued as it lacks the universal chemical interpretation of the ellipticity $\varepsilon$ e.g. double-bond $\varepsilon > 0.25$ vs. single bond character $\varepsilon \approx 0.10$. Other unsuitable scaling factor choices, on the basis of not attaining zero, included either ratios involving the $\lambda_1$ and $\lambda_2$ eigenvalue or any inclusion of the $\lambda_3$ eigenvalue. The $\lambda_3$ eigenvalue is unsuitable because it contains no information about the least ($\underline{e}_1$) and most ($\underline{e}_2$) preferred directions of the total charge density $\rho(\mathbf{r})$ accumulation.

With $n$ scaled eigenvector $\underline{e}_2$ tip path points $q_i = r_i + \varepsilon_i\underline{e}_{2,i}$ on the path $q$ where $\varepsilon_i$ = ellipticity at the $i^{th}$ bond-path point $r_i$ on the bond-path $r$. It should be noted that the bond-path is associated with the $\lambda_3$ eigenvalues of the $\underline{e}_3$ eigenvector does not take into account differences in the $\lambda_1$ and $\lambda_2$ eigenvalues of the $\underline{e}_1$ and $\underline{e}_2$ eigenvectors. Analogously, for the $\underline{e}_1$ tip path points we have $p_i = r_i + \varepsilon_i\underline{e}_{1,i}$ on the path $p$ where $\varepsilon_i$ = ellipticity at the $i^{th}$ bond-path point $r_i$ on the bond-path $r$.

We will refer to the next-generation QTAIM interpretation of the chemical bond as the *bond-path framework set* that will be denoted by $\mathbb{B}$, where $\mathbb{B} = \{p, q, r\}$. This effectively means that in the most general case a bond is comprised of three 'linkages'; $p, q$ and $r$ associated with the $\underline{e}_1$, $\underline{e}_2$ and $\underline{e}_3$ eigenvectors, respectively. The $p$ and $q$ parameters define eigenvector-following paths with lengths $\mathbb{H}^*$ and $\mathbb{H}$, see **Scheme 2**:

$$\mathbb{H}^* = \sum_{i=1}^{n-1}|p_{i+1} - p_i| \tag{3a}$$
$$\mathbb{H} = \sum_{i=1}^{n-1}|q_{i+1} - q_i| \tag{3b}$$

The lengths of the *eigenvector-following paths* $\mathbb{H}^*$ or $\mathbb{H}$ refers to the fact that the tips of the scaled $\underline{e}_1$ or $\underline{e}_2$ eigenvectors will sweep out along the extent of the bond-path, defined by the $\underline{e}_3$ eigenvector, between the two bonded nuclei that the bond-path connects. In the limit of vanishing ellipticity $\varepsilon = 0$, *for all* steps $i$ along the bond-path, one has $\mathbb{H}$ = BPL.

From the form of $p_i = r_i + \varepsilon_i\underline{e}_{1,i}$ and $q_i = r_i + \varepsilon_i\underline{e}_{2,i}$ we see for shared-shell *BCP*s, that in the limit of the ellipticity $\varepsilon \approx 0$ i.e. corresponding to single bonds, we then have $p_i = q_i = r_i$ and therefore the value of the lengths $\mathbb{H}^*$ and $\mathbb{H}$ attain their lowest limit; the bond-path length ($r$) BPL. Conversely, higher values of the ellipticity $\varepsilon$, for instance, corresponding to double bonds will always result in values of $\mathbb{H}^*$ and $\mathbb{H} >$ BPL.

For a non-torsional distortion such as a pure BLA coordinate motion there is no change in the orientation of the $\underline{e}_1$ and $\underline{e}_2$ eigenvectors of the $\{\underline{e}_1, \underline{e}_2, \underline{e}_3\}$ bond-path framework for all values of the scaling factor $\varepsilon_i$. A consequence of this is that the $\mathbb{H}^*$ and $\mathbb{H}$, that are constructed with the $\underline{e}_1$ and $\underline{e}_2$ eigenvectors respectively, will not correlate with the relative energy $\Delta E$ for a BLA distortion unlike an applied torsion $\theta$. Additionally, because $\mathbb{H}^*$ and $\mathbb{H}$ are defined by the distances swept out by the $\underline{e}_2$ tip path points $p_i = r_i + \varepsilon_i \underline{e}_{1,i}$ and $q_i = r_i + \varepsilon_i \underline{e}_{2,i}$ respectively and the scaling factor, $\varepsilon_i$ is identical in equation **(3a)** and equation **(3b)** therefore for a linear bond-path $r$ then $\mathbb{H}^* = \mathbb{H}$. The bond-path framework set $\mathbb{B} = \{p, q, r\}$ should consider the bond-path to comprise the *unique* paths, $p$, $q$ and $r$, swept out by the $\underline{e}_1$, $\underline{e}_2$ and $\underline{e}_3$, eigenvectors that form the eigenvector-following paths with lengths $\mathbb{H}^*$, $\mathbb{H}$ and BPL respectively. The paths $p$ and $q$ are unique even when the lengths of $\mathbb{H}^*$ and $\mathbb{H}$ are the same or very similar because $p$ and $q$ traverse different regions of space. Bond-paths $r$ with non-zero bond-path curvature which will result in $\mathbb{H}^*$ and $\mathbb{H}$ with different values, this is more likely to occur for the equilibrium geometries of closed-shell *BCP*s than for shared-shell *BCP*s. This is because the $p$ and $q$ paths will be different because of the greater distance travelled around the outside of a twisted bond-path $r$ compared with the inside of the same twisted bond-path $r$.

Analogous to the bond-path curvature, see equation **(2)**, we may define dimensionless, *fractional* versions of the eigenvector-following path with length $\mathbb{H}$ where several forms are possible and not limited to the following:

$\mathbb{H}_f = (\mathbb{H} - BPL)/BPL$               **(4a)**

$\mathbb{H}_{f\theta min} = (\mathbb{H} - \mathbb{H}_{\theta min})/\mathbb{H}_{\theta min}$,             **(4b)**

where $\mathbb{H}_{\theta min}$ is the length swept out by the scaled $\underline{e}_2$ eigenvectors using the value of the torsion $\theta$ at the energy minimum. Similar expressions for $\mathbb{H}^*_f$ and $\mathbb{H}^*_{f\theta min}$ can be derived using the $\underline{e}_1$ eigenvectors.

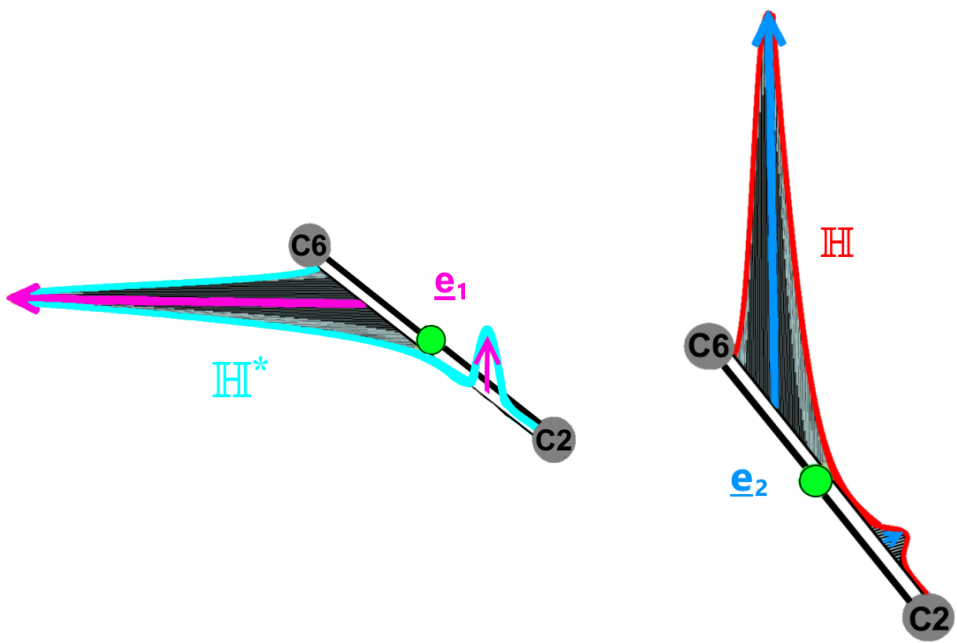

**Scheme 2.** The pale blue line in sub-figure **(a)** represents the path, referred to as the eigenvector-following path with length $\mathbb{H}^*$, swept out by the tips of the scaled $\underline{e}_1$ eigenvectors, shown in magenta, and defined by equation **(2a)**. The red path in sub-figure **(b)** corresponds to the eigenvector-following path with length $\mathbb{H}$, constructed from the path swept out by the tips of the scaled $\underline{e}_2$ eigenvectors, shown in mid blue and is defined by equation **(2b)**. The pale and mid blue arrows representing the $\underline{e}_1$ and $\underline{e}_2$ eigenvectors are scaled by the ellipticity ε respectively, where the vertical scales are exaggerated for visualization purposes. The green sphere indicates the position of a given *BCP*. Details of how to implement the calculation of the eigenvector-following paths with lengths $\mathbb{H}^*$ and $\mathbb{H}$ are provided in the **Supplementary Materials S10**.

The form of $\mathbb{H}_f$ defined by equation **(4a)** is the closest to the spirit of the bond-path curvature, equation **(2)**.

A bond within QTAIM is defined as being the bond-path traversed along the $\underline{e}_3$ eigenvector of the $\lambda_3$ eigenvalue from the bond-path, but, as a consequence of equation **(3)**, this definition should be expanded. This next-generation QTAIM definition of a bond should consider the bond-path to comprise the two paths swept out by the $\underline{e}_1$ and $\underline{e}_2$ eigenvectors that form the eigenvector-following path with length $\mathbb{H}^*$ and $\mathbb{H}$, respectively. For the most general case of a bond-path *r* with non-zero curvature we can have a bond-path framework set $\mathbb{B} = \{(p_0,p_1), (q_0,q_1), r\}$ as the set of five unique paths through the 3-D Cartesian space, where the subscripts "0" and "1" refer to the ground state $S_0$ and the first excited state $S_1$ respectively. Conversely, for a bond-path *r* with zero bond-path curvature we have $\mathbb{H}^* = \mathbb{H}$ therefore $p_0 = q_0$ and $p_1 = q_1$ and $\mathbb{B}$ reduces to three unique elements $\mathbb{B} = \{(p_0,p_1), r\}$.

## 3. Computational Details

The potential energy surfaces (PESs) of the $S_1$ and $S_0$ states of fulvene along the bond stretching and double bond torsion coordinates were investigated using the state-interaction state-averaged REKS (SI-SA-REKS) method[18]. The SI-SA-REKS method, denoted further on as SSR, for brevity, employs ensemble density functional theory (eDFT) to describe the strong non-dynamic electron correlation, *e.g.*, resulting from breaking of chemical bonds, and to provide variational description of the excited states similar, in its spirit, to state-averaged CASSCF (SA-CASSCF) methodology. By contrast to SA-CASSCF, the SSR method includes the dynamic electron correlation through the use of approximate density functionals; hence capable of delivering results matching the accuracy of high-level multi-reference wavefunction theory methods, such as, *e.g.*, multi-reference configurational interaction (MRCI), when describing the PESs of ground and excited states of molecules and conical intersections between the PESs[19,20].

In this work, the SSR method is employed in connection with the ωPBEh range-separated hybrid density functional[21] and the cc-pVDZ basis set[22], the SSR-ωPBEh/cc-pVDZ method. The geometries of the ground and excited states minima and the minimum energy pathways connecting the critical points were optimized using the analytical energy derivatives as described in Ref [23]. The computations were carried out using the beta-testing version of the TeraChem® program (v1.92P, release 7f19a3bb8334)[24–29]. Optimization of conical intersection geometries were carried out using the penalty function method with the analytical energy gradients as implemented in the CIOpt program[30]. Optimization of the geometries along the MEPs were carried out using the nudged elastic band (NEB) formalism[31] with fixed end points. The MEPs comprise 40 points between the Franck-Condon (FC) geometry and the planar $C_{2v}$ symmetric conical intersection ($CI_{plan}$) and the torsional $CI_{tor}$. At all the points along the MEPs the relaxed density matrices for the $S_0$ and the $S_1$ states[23] were calculated and analyzed using the AIMALL software suite[32].

## 4. Results and discussions

*4.1 The excited state deactivation reaction through the both the planar and torsional CIs*

The $S_0$ state of fulvene near the equilibrium conformation (FC) is best described by the Lewis structure shown in **Scheme 3**. The ground state electronic structure at the FC geometry is characterized by considerable conjugation of the π-bonds of fulvene, which leads to shortening of the (formally) single bonds. Excitation to the $S_1$ state results in breaking of the exocyclic π-bond, see the Lewis structures in **Scheme 3**, and formation of a diradical state. The calculated vertical excitation energy at the FC geometry is 3.65 eV, in a reasonable agreement with the experimental gas phase vertical excitation energy of 3.44 eV[33].

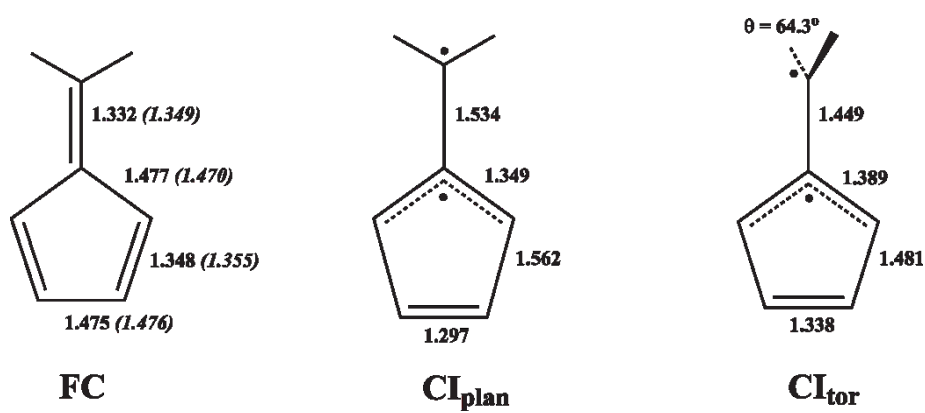

**Scheme 3**. Geometries of the FC point ($S_0$ state) and the two conical intersections, $CI_{plan}$ and $CI_{tor}$, optimized in this work. Key geometric parameters (in Å) are shown. In parentheses, the gas phase experimental data from[34]. Lewis structures of the $S_0$ (FC) and $S_1$ ($CI_{plan}$ and $CI_{tor}$) states are shown.

Crossing between the $S_0$ and $S_1$ states becomes possible due to geometric distortions, which destabilize the $S_0$ state. As in the $S_1$ state the exocyclic π-bond is broken, stretching and/or torsion of this bond do not cost much energy thus leading to only minor variation of the $S_1$ energy. For the $S_0$ state however, these distortions result in a considerable destabilization and reduction of the $S_1/S_0$ gap. With the use of the penalty function method, two conical intersections were identified and optimized in this work. The planar $C_{2v}$ symmetric CI$_{plan}$ features stretched exocyclic C-C bond as well as inverted lengths of the previously (formally) single and double bonds, see **Scheme 3**. This distortion is known as the bond-length-alternation (BLA) and is calculated as the difference between the average length of the formally single bonds and the average length of the formally double bonds in the molecule. The MEP connecting the FC point with CI$_{plan}$, see **Figure 1(a)**, shows that the BLA drops from a positive value of 0.13 Å (indicating that the single bonds are longer that the doubles) to a negative value of -0.22 Å (single bonds shorter than doubles). With respect to the S$_0$ FC energy, CI$_{plan}$ lies slightly higher (88.0 kcal/mol) than the FC S$_1$ point (84.2 kcal/mol), see **Figure 1(a)**.

In the CI$_{tor}$ structure, destabilization of the S$_0$ energy is achieved by combined BLA and torsional distortions. The degree of bond stretching/bond contraction is not as pronounced as in the CI$_{plan}$ structure; the BLA decreases to the values in the range of (-0.02, -0.09) Å, see **Figure 1(b)**. The FC-CI$_{tor}$ MEP in **Figure 1(b)** shows that the BLA distortion rapidly decreases at the beginning of the path, while the torsion θ about the exocyclic C-C bond picks up somewhat later along the MEP and reaches torsional angle of 64.3º. The optimized FC-CI$_{tor}$ MEP is generally consistent with the shape of the S$_0$ and S$_1$ PESs evaluated on the basis of the CASSCF calculations[35]. The CI$_{tor}$ occurs at 76.3 kcal/mol w.r.t. the S$_0$ FC energy and is 11.7 kcal/mol lower than CI$_{plan}$. The energy gap between the CI points obtained here is in agreement with the CASSCF value of 9.7 kcal/mol from Ref[36].

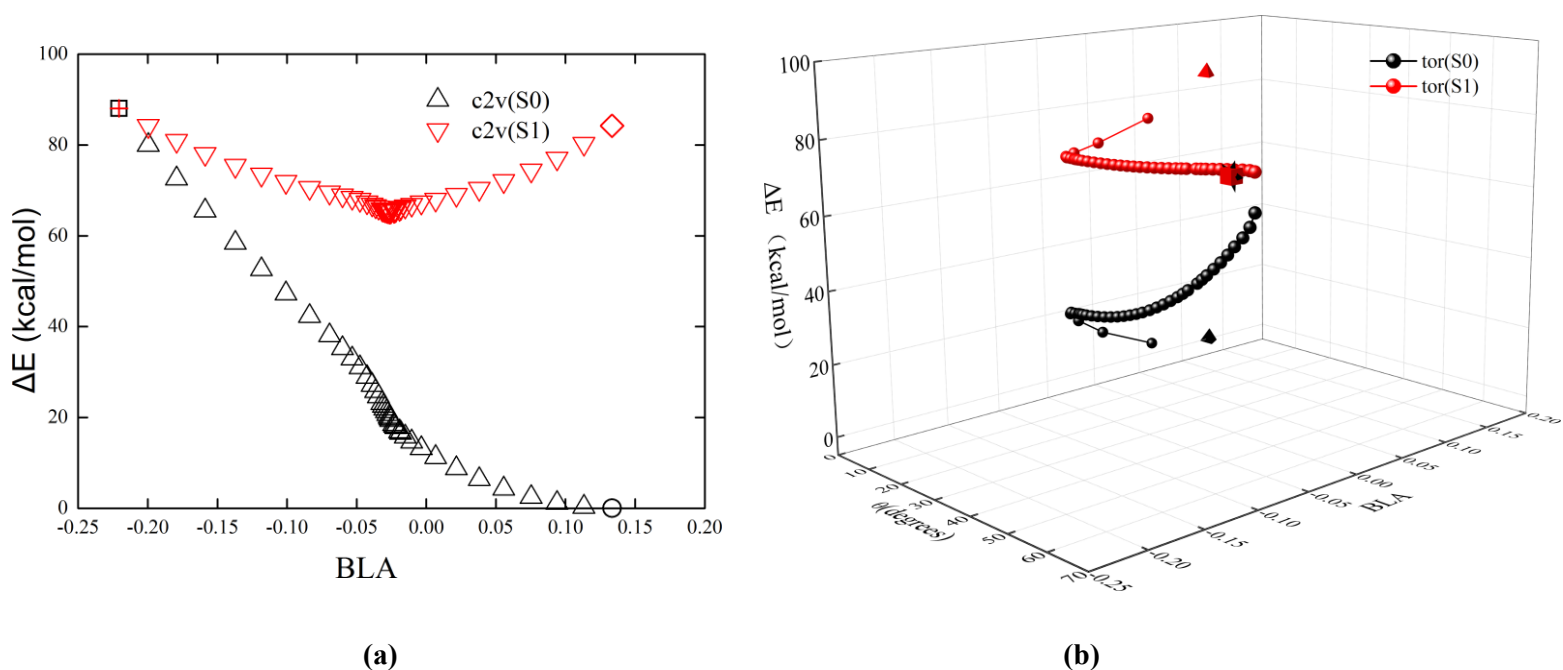

(a)  (b)

**Figure 1.** The profiles of the S$_1$ (red) and S$_0$ (black) PESs along the MEPs connecting the FC point with the planar C$_{2v}$ (a) and torsional CIs **in (b)**. The relative energy ΔE (in kcal/mol) is given with respect to the energy of the S$_0$ minimum. For the FC-CI_plan MEP, variation of the energy is shown with respect to the BLA distortion (in Å). For the FC-CI_tor MEP, a 3-D graph in sub-figure **(b)** shows the relative energies ΔE as functions of both the BLA distortion and the torsional angle θ (in degrees). The FC point in sub-figure **(a)** is given by a black circle (S$_0$) and by a red diamond (S$_1$); the CI_plan is shown by a black square and red cross, respectively. In sub-figure **(b)**, the FC point is shown by black (S$_0$) and red (S$_1$) pyramids and the CI_tor is shown by a black star and red cube, respectively.

*4.2 QTAIM BCP and bond-path analysis of the deactivation reaction*

First we consider variation of the C2-C6 *BCP* in the deactivation reaction of fulvene along the FC-to-CI$_{plan}$ MEP, see **Scheme 1 and Scheme 3** and **Figure 1-3**.

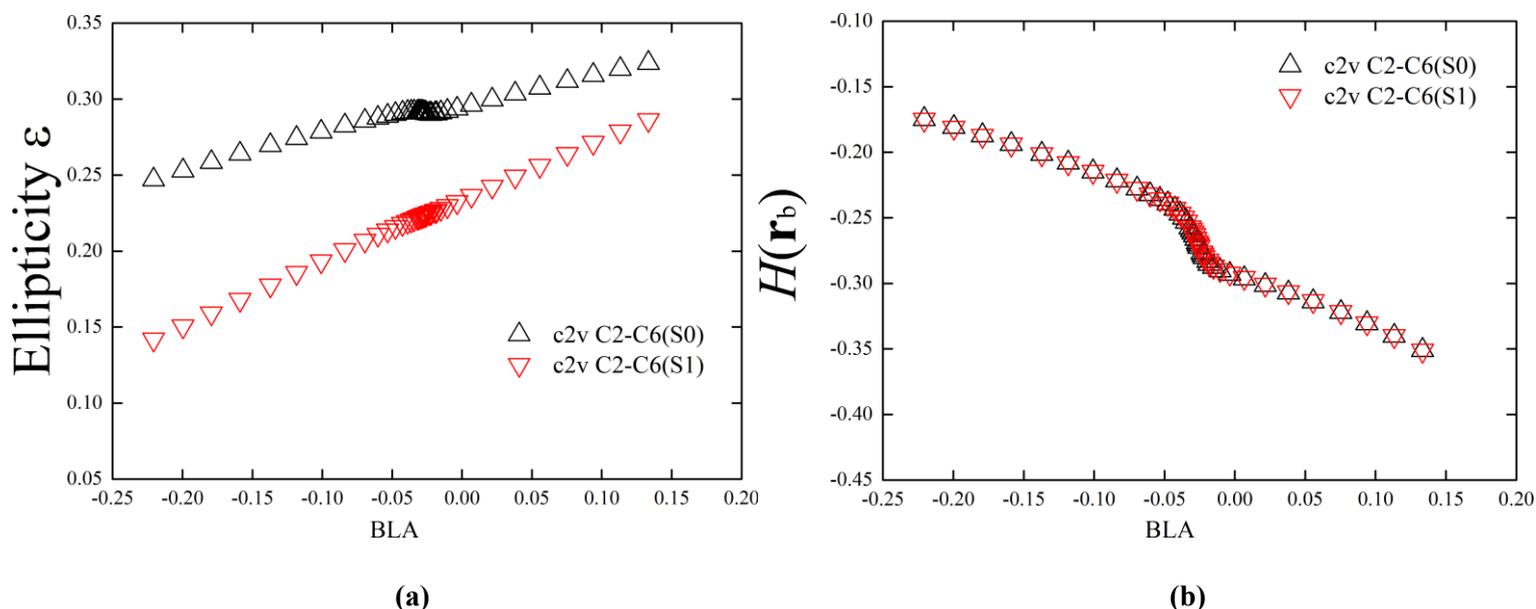

(a)  (b)

**Figure 2.** The variation of the ellipticity ε, total local energy density $H(r_b)$, with the BLA coordinate of the C2-C6 *BCP* are presented in sub-figures **(a)** and **(b)** respectively. See the caption of **Figure 1** for further details.

Along the FC-to_CI$_{plan}$ MEP, the value of the ellipticity ε of the C2-C6 *BCP* in the $S_1$ electronic state varies monotonically from a value typical for a double bond (right side of **Figure 2(a)**) to a value characteristic of a single bond (at the CI$_{plan}$, left side of **Figure 2(a)**). The corresponding values of the ellipticity ε of the C2-C6 *BCP* for the $S_0$ state indicate retaining the double bond character for the entire MEP. Consistent with the ellipticity ε, the $H(r_b)$ values of the C2-C6 *BCP* increase along the MEP, however, the $H(r_b)$ does not distinguish between the $S_0$ and $S_1$ states, which limits the usefulness of this QTAIM measure, see **Figure 2(b)**.

The values of the stress tensor $\lambda_{3\sigma}$ are always higher for the $S_0$ state compared with the $S_1$ state. This indicates the greater stability of the C2-C6 *BCP* in the $S_0$ state compared with the $S_1$ state and is consistent with the relative energy ΔE, see the **Supplementary Materials S8** and **Figure 1(a)** respectively. The stress tensor stiffness $\mathbb{S}_\sigma$ increases along the reaction coordinate, which is consistent with both the ellipticity ε and the total local energy density $H(r_b)$, see the **Supplementary Materials S8**. There is a slight increase in the $\mathbb{S}_\sigma$ values for the $S_1$ state relative to the $S_0$ state values, which is not revealed from the other scalar values.

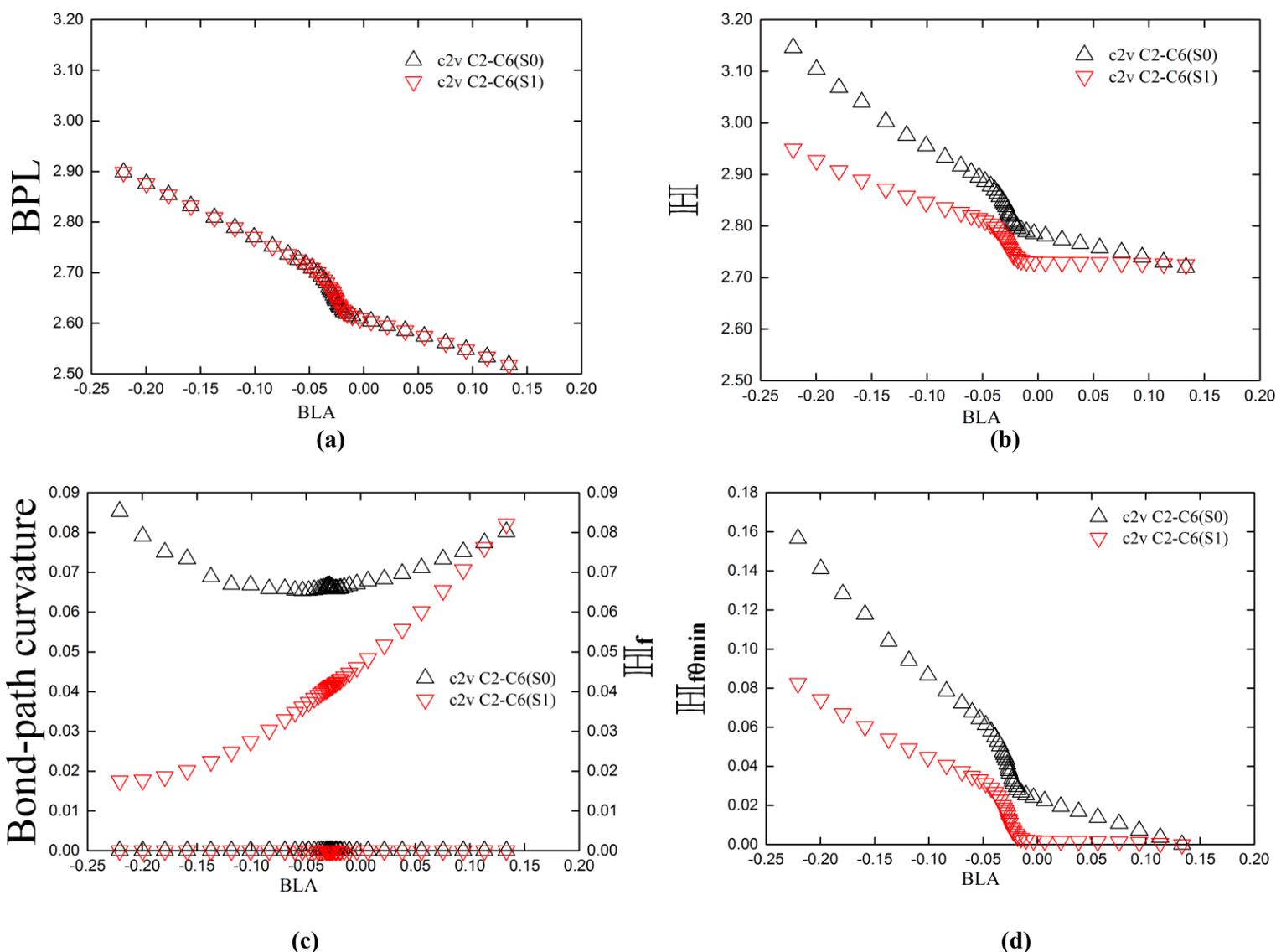

**Figure 3.** The variation of the bond-path length (BPL) of the torsional C2-C6 *BCP* with the BLA is shown in sub-figure **(a)**. The corresponding variations of length $\mathbb{H}$ of the eigenvector-following path and the corresponding fractional length $\mathbb{H}_f = (\mathbb{H} - \text{BPL})/\text{BPL}$ are shown in sub-figure **(b)** and sub-figure **(c)** respectively, where the bond-path curvature, showing zero variability, is included in sub-figure **(c)** for comparison. The corresponding variation of the fractional length $\mathbb{H}_{f\theta min} = (\mathbb{H} - \mathbb{H}_{\theta min})/\mathbb{H}_{\theta min}$ where $\theta_{min}$ refers to the value of $\theta$ at the energy minimum, is shown in sub-figure **(d)**, see the caption of **Figure 1** for further details.

The variation of the BPL with the BLA coordinate is consistent with the variation of the $H(\mathbf{r_b})$ values in that the BPL decreases with the increase in the BLA coordinate showing the weakening of the C2-C6 *BCP*, see **Figure 3(a)**. The decrease of BPL with BLA coordinate is also consistent with the increase in ellipticity ε, compare **Figure 3(a)** with **Figure 2(a)**. The negative values of the BLA coordinate correspond to the occurrence of a bond-length inversion.

The variation of the BPL with the BLA coordinate does not differentiate between the $S_0$ and $S_1$ states, this lack of differentiation between the $S_0$ and $S_1$ states also occurs for $H(\mathbf{r_b})$. The variation of the bond-path curvature with the BLA coordinate is insignificant, showing the limitation of this QTAIM measure for the purpose of the C2-C6 *BCP* deactivation reaction, see **Figure 3(b)**.

The length $\mathbb{H}$ of the eigenvector-following path constructed using the $\underline{\mathbf{e}}_2$ eigenvector shown in equation **4(a)**, varies significantly along the C2-C6 *BCP* bond-path and more so than the BPL. Additionally, the $\mathbb{H}$ values differentiate between the $S_0$ and $S_1$ electronic states, see **Figure 3(b)**. The variation of the length $\mathbb{H}^*$ of the eigenvector-following path, constructed using the $\underline{e}_1$ eigenvector shown in equation **4(b)**, along the BLA coordinate, is indistinguishable from the variation of $\mathbb{H}$ with BLA due to the bond-path (*r*) possessing almost no bond-path curvature, see **Figure 3(c)**, see the text surrounding equation **4(a)** and the theory section 2.2 for further explanation and the **Supplementary Materials S9**. The fractional eigenvector-following path with lengths $\mathbb{H}_f$ and $\mathbb{H}_{f\theta min}$, unlike the bond-path curvature, vary significantly along the BLA coordinate, see **Figure 3(c-d)** respectively.

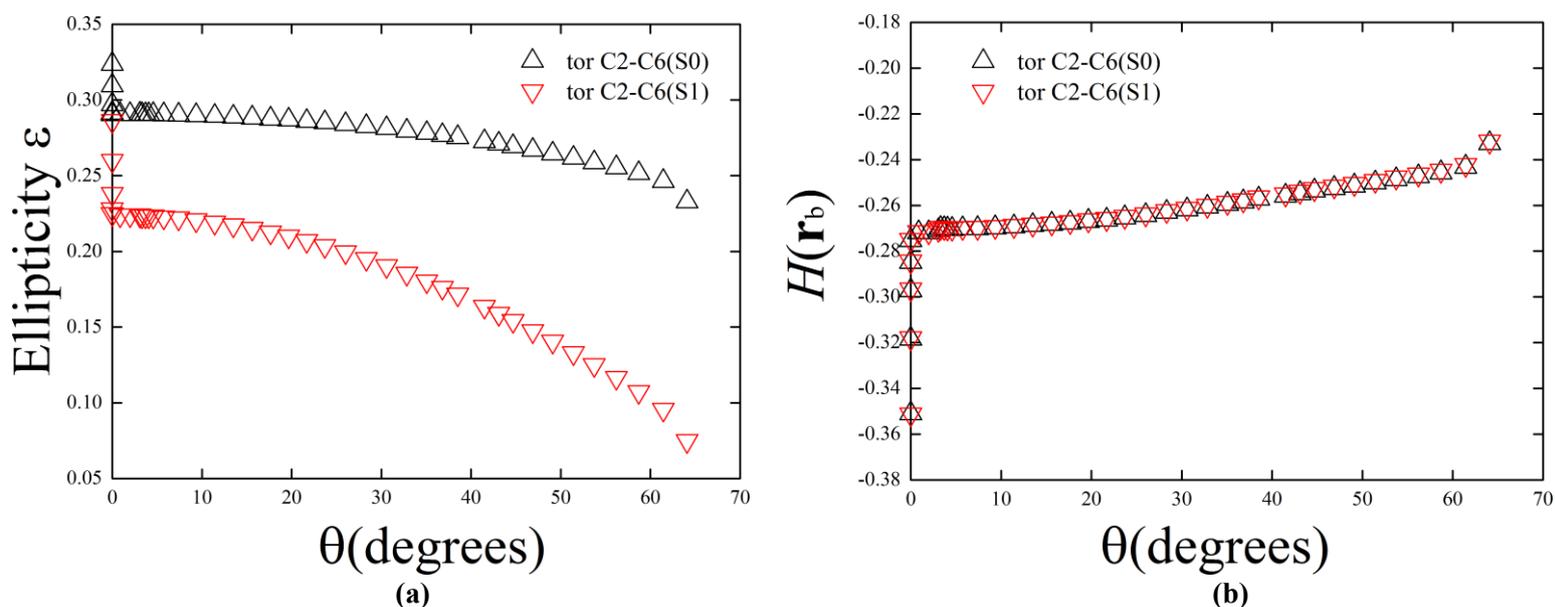

**Figure 4.** The variation of the ellipticity ε, total local energy density $H(\mathbf{r_b})$, with the torsion θ of the C2-C6 *BCP* are presented in sub-figures **(a)** and **(b)** respectively. See the caption of **Figure 1** for further details.

Along the torsional FC-to-CI$_{tor}$ MEP, the ellipticity ε of the C2-C6 *BCP* for the $S_0$ state indicates double bond character (ε > 0.20), at θ = 64° whilst for the $S_1$ the C2-C6 *BCP* has single bond character indicating the deactivation reaction is facilitated by the $S_1$ excited state, see **Figure 4(a)**. Similar to the stretching MEP, the total local energy density $H(\mathbf{r_b})$ does not differentiate between the $S_0$ and $S_1$ states, although the result of the monotonically increasing $H(\mathbf{r_b})$ values is consistent with trend in the ellipticity ε values decreasing for both the $S_0$ and $S_1$ states. The stress tensor $\lambda_{3\sigma}$ is always more positive for the $S_0$ state compared with the $S_1$ state, indicating the greater stability of the C2-C6 *BCP* in the $S_0$ state and is consistent with the relative energy ΔE, see **Figure S7(f) of the Supplementary Materials S7** and **Figure 1(b)** respectively. The lower values of the stress tensor stiffness $\mathbb{S}_\sigma$ indicate that in the $S_0$ state the C2-C6 *BCP* more readily undergoes a torsion θ deformation than in the $S_1$ state, see **Figure S8(f)** of the **Supplementary Materials S8**.

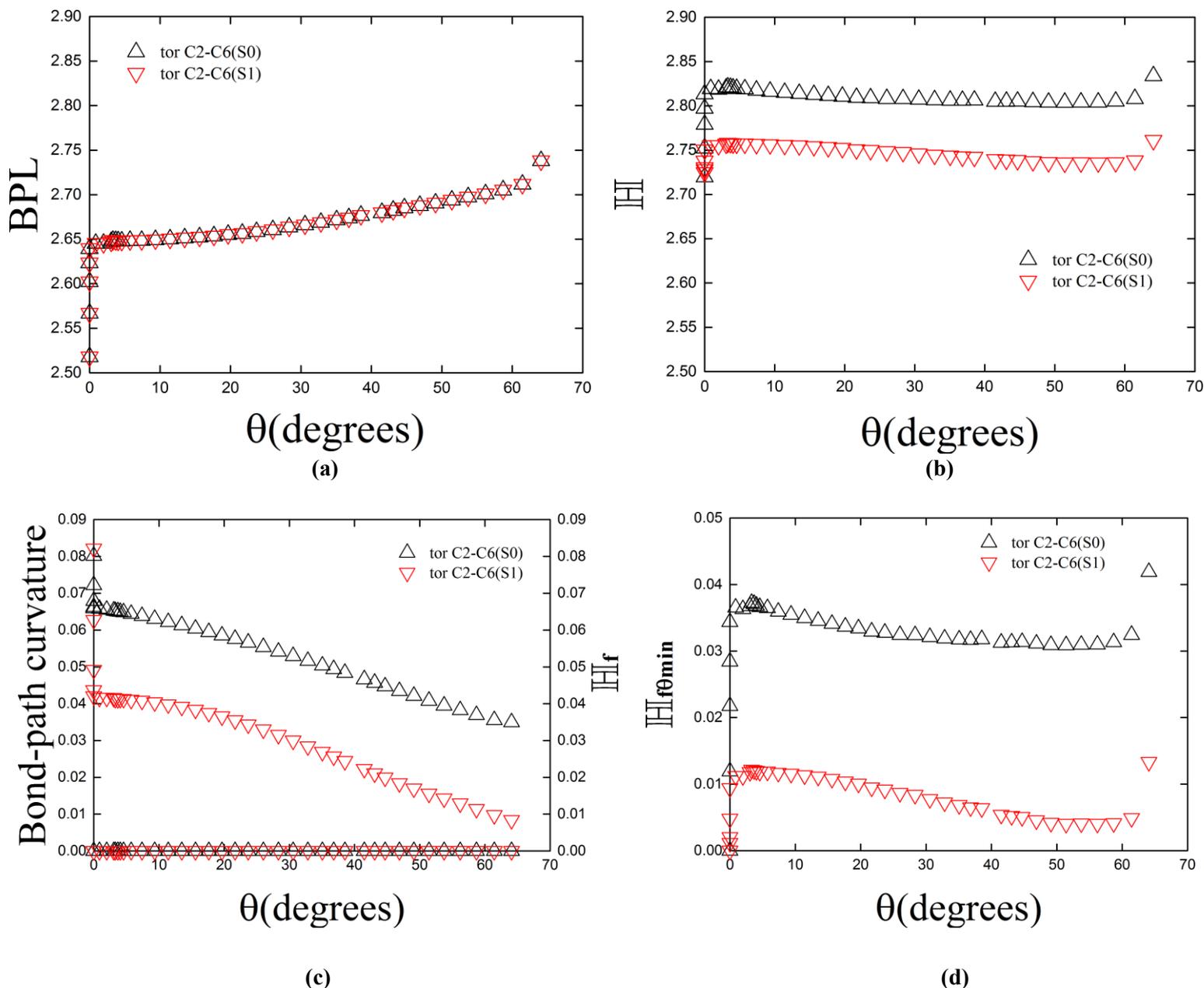

**Figure 5.** The variation of the bond-path length (BPL) of the torsional C2-C6 *BCP* with torsion θ is shown in sub-figure **(a)**. The corresponding variations of $\mathbb{H}$ and the fractional length $\mathbb{H}_f$ are shown in sub-figure **(b)** and sub-figure **(c)** respectively, where the bond-path curvature, showing zero variability, is included in sub-figure **(c)** for comparison. The corresponding variation of the fractional length $\mathbb{H}_{f\theta min}$, is shown in sub-figure **(d)**, see the caption of **Figure 3** for further details.

The BPL along the torsion θ does not differentiate between the $S_0$ and $S_1$ states and the variation of the bond-path curvature with torsion θ is insignificant, see **Figure 5(a)** and **Figure 5(c)** respectively. Both of these findings were also present for the reaction along the BLA coordinate, see **Figure 3(a)** and **Figure 3(c)** respectively.

There are significant differences between the lengths $\mathbb{H}$ of the $S_0$ and $S_1$ states for the variation with the torsion θ as was the case for the BLA coordinate, see **Figure 5(b)** and **Figure 3(b)**.

The eigenvector-following path with length $\mathbb{H}$, see **Figure 5(b)**, is shorter for the $S_1$ compared with the $S_0$ since this corresponds to lower values of the ellipticity ε, see **Figure 4(a)**, indicating that the torsion θ is easier to perform in the excited state from the form of $q_i = r_i + \varepsilon_i \underline{e}_{2,i}$. This finding is consistent with our earlier work with the value of the ellipticity ε of the torsion *BCP* being lowered by photo-isomerization[6,37,38]. The same trend is also true for the fractional lengths $\mathbb{H}_f$ and $\mathbb{H}_{f\theta min}$ for the BLA and torsion θ, see **Figure 3(c-d)** and **Figure 5(c-d)** respectively. The lengths $\mathbb{H}$ and $\mathbb{H}^*$ are almost degenerate for values of the torsion θ ≈ 35°, as was found for the variation with the BLA coordinate: this is due to the negligible bond-path (*r*) curvature, see the **Supplementary Materials S9**.

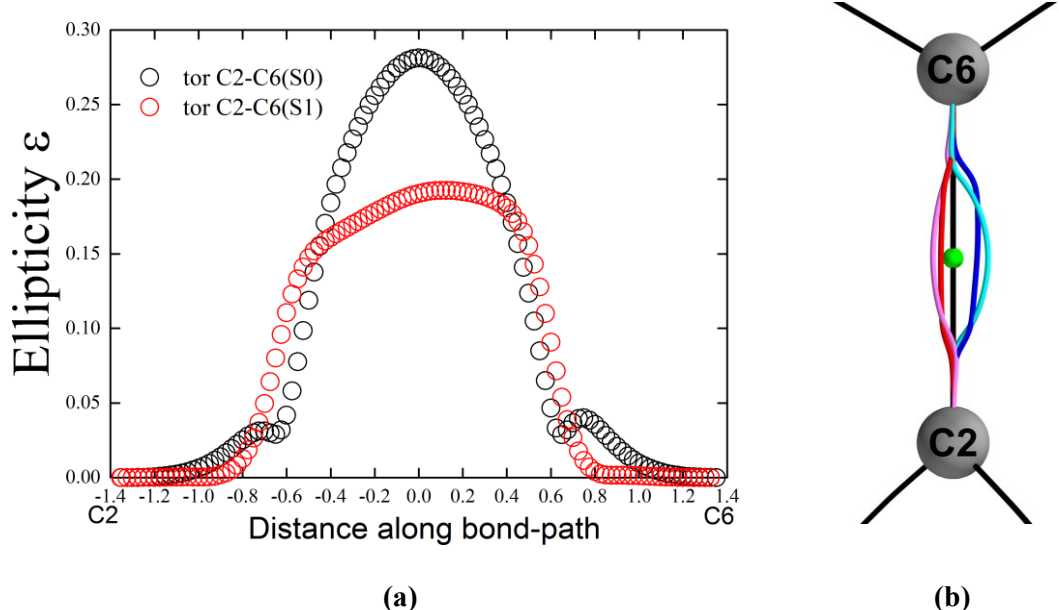

(a)            (b)

**Figure 6.** The $S_0$ and $S_1$ profiles of the scaling factor for the construction of the *p* and *q* paths, the ellipticity ε, along the torsion C2-C6 *BCP* bond-path for the fulvene molecular graph subjection to a torsion θ, where θ = 30.0° is presented in sub-figure **(a)**, the ellipticity ε is the scaling factor used in the construction of the *p* and *q* paths where $p_i = r_i + \varepsilon_i \underline{e}_{1,i}$ and $q_i = r_i + \varepsilon_i \underline{e}_{2,i}$ with lengths $\mathbb{H}^*$ and $\mathbb{H}$ respectively. The corresponding five calculated unique paths that comprise the bond-path framework set $\mathbb{B}$; $p_0$ (pale-blue), $q_0$ (pink), $p_1$ (mid-blue), $q_1$ (red) and the bond-path *r* (black) where the undecorated green sphere indicates the position of torsion C2-C6 *BCP*.

The $S_0$ and $S_1$ plots of profile of the scaling factor, the ellipticity ε, along the torsion C2-C6 *BCP* bond-path for a torsion θ, where θ = 30.0°, is presented in **Figure 6(a)**. The plot demonstrates differences in the scaling factor for the $S_0$ and $S_1$ along the torsion C2-C6 *BCP* bond-path, of particular note are unexpected features including significant asymmetry in the $S_1$ profile and also local maxima in the $S_0$ profile at approximately -0.7 a.u. and 0.75 a.u.. The corresponding frame-path set $\mathbb{B} = \{(p_0,p_1), (q_0,q_1), r\}$ is presented and demonstrates significant variations between the paths corresponding to the least $p_0$ (pink), $p_1$ (red) and most $q_0$ (pink), $q_1$ (red) preferred directions, see **Figure 6(b)**.

# 5. Conclusions

We have introduced a next-generation QTAIM-based vector representation of the chemical bond; the bond-path framework set $\mathbb{B} = \{p, q, r\}$ with the $p$, $q$ and $r$ being associated with each of the $\underline{\mathbf{e}}_1$, $\underline{\mathbf{e}}_2$ and $\underline{\mathbf{e}}_3$ eigenvectors of the Hessian of $\rho(\mathbf{r})$ respectively. The scalar lengths of $p$, $q$ and $r$ correspond to the eigenvector path lengths $\mathbb{H}^*$ and $\mathbb{H}$ and the bond-path length (BPL) respectively.

The purpose of the bond-path framework set $\mathbb{B}$ is to provide a directional representation of the chemical bond, for instance, where $\mathbb{H}^*$ and $\mathbb{H}$ are the vector-based counterpart to the scalar ellipticity ε and can be used for the interpretation of excited state reactions. This was undertaken by quantifying the least ($\underline{\mathbf{e}}_1$) and most preferred ($\underline{\mathbf{e}}_2$) directions of the accumulation of $\rho(\mathbf{r})$ in terms of the eigenvector-following paths $p$ and $q$ with lengths $\mathbb{H}^*$ and $\mathbb{H}$, respectively. For shared-shell *BCP*s, the value of $\mathbb{H}^*$ and $\mathbb{H}$ attain their lowest limit; that of the bond-path length (BPL) for values of the ellipticity ε that correspond to single bonds. Conversely, higher values of the ellipticity ε corresponding to double bonds will always result in values of $\mathbb{H}^*$ and $\mathbb{H} >$ BPL. Therefore, greater values of $\mathbb{H}^*$ and $\mathbb{H} >$ BPL indicate a greater resistance to an applied torsion θ and the minimum value of resistance to torsion can be occur for $\mathbb{H}^* = \mathbb{H} =$ BPL.

The familiar QTAIM bond-path ($r$) was found to be unable to distinguish between the $S_0$ and $S_1$ electronic states, however, the complete set comprising $\mathbb{B} = \{p, q, r\}$ provides an orthogonal framework to quantify any type of bond distortion. For instance, torsion θ, can be quantified using $p$ and $q$ (constructed from $\underline{\mathbf{e}}_1$ and $\underline{\mathbf{e}}_2$) and BLA using $r$ (constructed from $\underline{\mathbf{e}}_3$) as well as any bond motion that can occur in general reactions.

We have applied $\mathbb{B}$, to a fulvene molecule undergoing excited state deactivation through conical intersections corresponding to bond stretching (BLA distortion) and double bond torsion θ. We find that the values of the lengths $\mathbb{H}$ and $\mathbb{H}^*$ associated with $p$ and $q$ respectively are almost degenerate for both the BLA distortion and torsion θ due to the presence in each case of negligible bond-path ($r$) curvature. The length $\mathbb{H}$ was able to distinguish between the $S_0$ and $S_1$ electronic states for both the excited state deactivation BLA and excited state deactivation torsional θ paths, unlike the bond-path ($r$) length.

To summarize, we now understand that the chemical bond in the ground state $S_0$, within the QTAIM framework, now comprises and additional two unique strands or paths; $p_0$ and $q_0$: the first excited state $S_1$ produces another two paths $p_1$ and $q_1$. Further higher excited states, $S_2$, $S_3$, …, $S_n$, will each yield a pair of paths $p_2$, $q_2$, $p_3$, $q_3$, …, $p_n$, $q_n$ where each $p_n$ and $q_n$ path has a defined length $\mathbb{H}_n$, $\mathbb{H}^*_n$ respectively. The morphology of the $p_n$ and $q_n$ paths defines the response to, for instance, photo-excitation, torsion θ, normal mode of vibration, in terms of the least ($\underline{\mathbf{e}}_1$) and most ($\underline{\mathbf{e}}_2$) preferred directions of the total electronic charge density $\rho(\mathbf{r})$ distribution.

**Acknowledgements**

The National Natural Science Foundation of China is gratefully acknowledged, project approval number: 21673071. The One Hundred Talents Foundation of Hunan Province and the aid program for the Science and Technology Innovative Research Team in Higher Educational Institutions of Hunan Province are also gratefully acknowledged for the support of S.J. and S.R.K.

# SUPPLEMENTARY MATERIALS

# Next-Generation Quantum Theory of Atoms in Molecules for the Ground and Excited States of Fulvene

Wei Jie Huang[1], Roya Momen[1], Alireza Azizi[1], Tianlv Xu[1], Steven R. Kirk[1*], Michael Filatov[1,2] and Samantha Jenkins[1*]

*Key Laboratory of Chemical Biology and Traditional Chinese Medicine Research and Key Laboratory of Resource Fine-Processing and Advanced Materials of Hunan Province of MOE, College of Chemistry and Chemical Engineering, Hunan Normal University, Changsha, Hunan 410081, China*

email: steven.kirk@cantab.net
email: mike.filatov@gmail.com
email: samanthajsuman@gmail.com

1. **Supplementary Materials S1.** The variation of the ellipticity ε of the C-C *BCPs* with BLA.

2. **Supplementary Materials S2.** The variation of the $H(\mathbf{r_b})$ of the C-C *BCPs* with BLA.

3. **Supplementary Materials S3.** The variation of the stress tensor eigenvalue $\lambda_{3\sigma}$ of the C-C *BCPs* with BLA.

4. **Supplementary Materials S4.** The variation of the stress tensor stiffness $\mathbb{S}_\sigma$ of the C-C *BCPs* with BLA.

5. **Supplementary Materials S5.** The variation of the ellipticity ε of the C-C *BCPs* with the torsion θ.

6. **Supplementary Materials S6.** The variation of the $H(\mathbf{r_b})$ of the C-C *BCPs* with the torsion θ.

7. **Supplementary Materials S7.** The variation of the stress tensor eigenvalue $\lambda_{3\sigma}$ of the C-C *BCPs* with the torsion θ.

8. **Supplementary Materials S8.** The variation of the stress tensor stiffness $\mathbb{S}_\sigma$ of the C-C *BCPs* with the torsion θ.

9. **Supplementary Materials S9.** The variation of the eigenvector-following path lengths $\mathbb{H}^*$ of the C2-C6 *BCP* with the BLA and torsion θ.

10. **Supplementary Materials S10.** Implementation details of the calculation of the eigenvector-following path lengths $\mathbb{H}$ and $\mathbb{H}^*$.

1. **Supplementary Materials S1.**

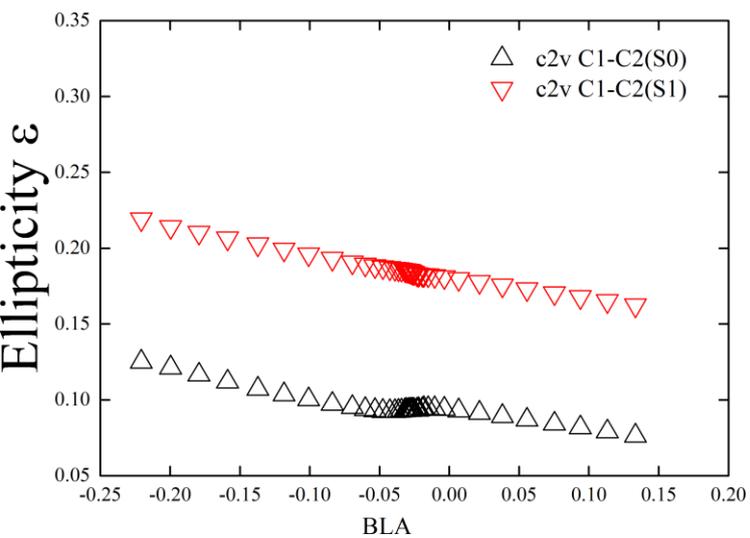

(a)

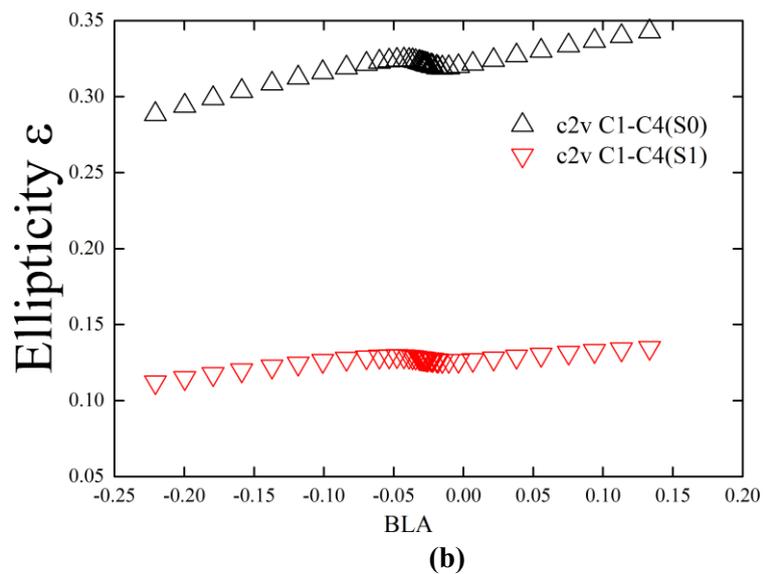

(b)

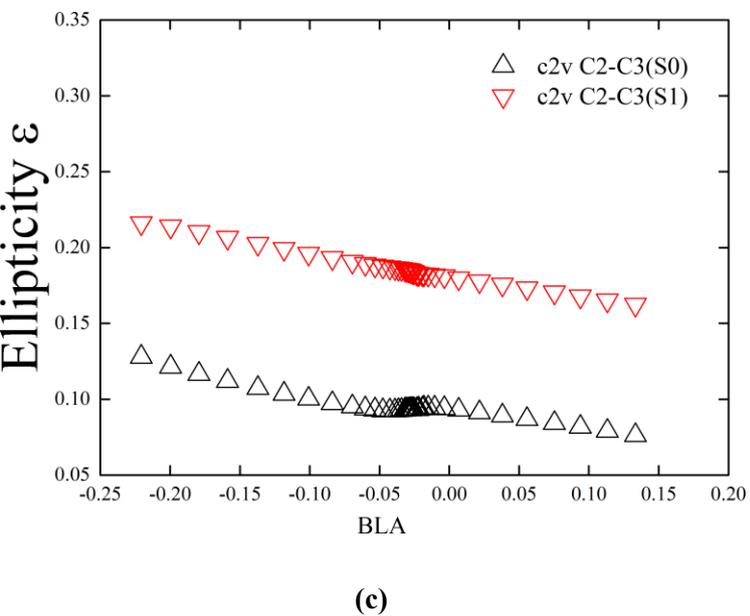

(c)

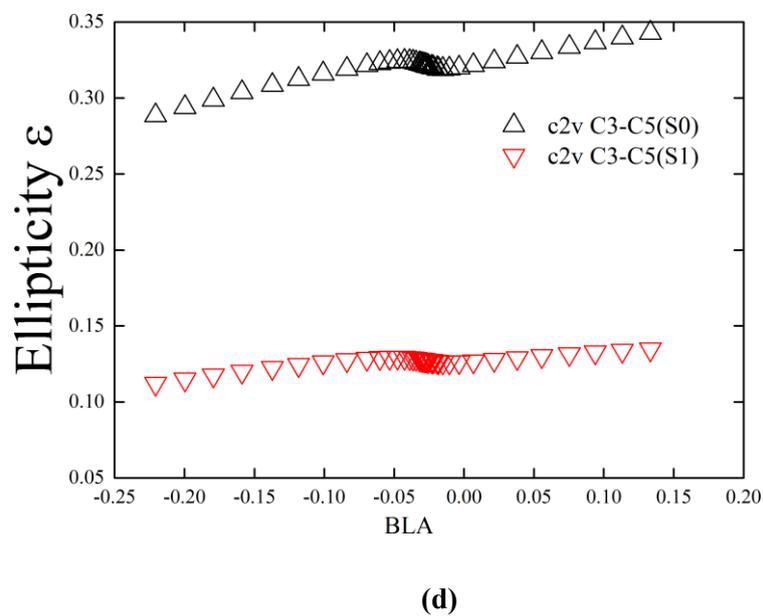

(d)

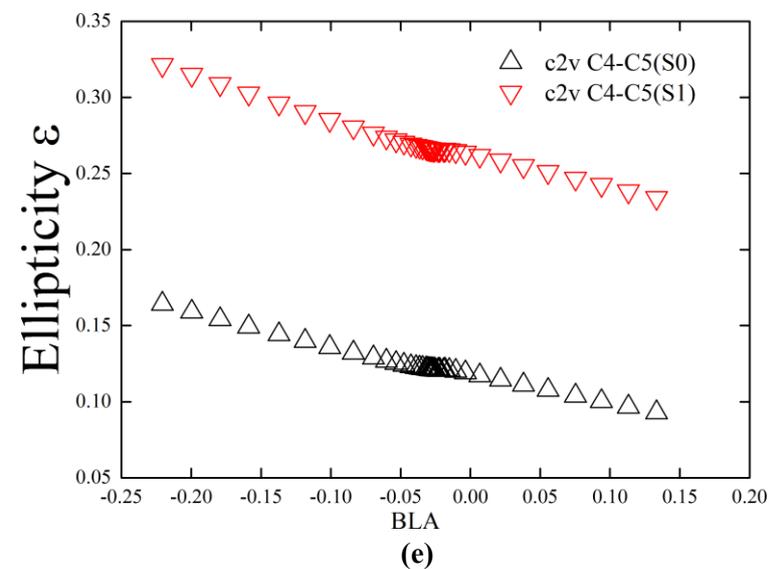

(e)

**Figure S1.** The variation of the ellipticity ε with the BLA for C-C *BCPs* is shown in sub-figures **(a)** - **(e)**.

## 2. Supplementary Materials S2.

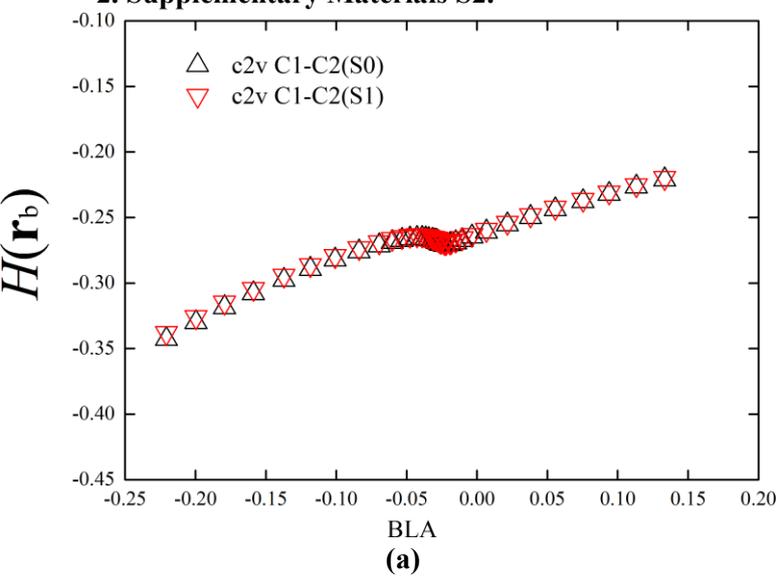

(a)

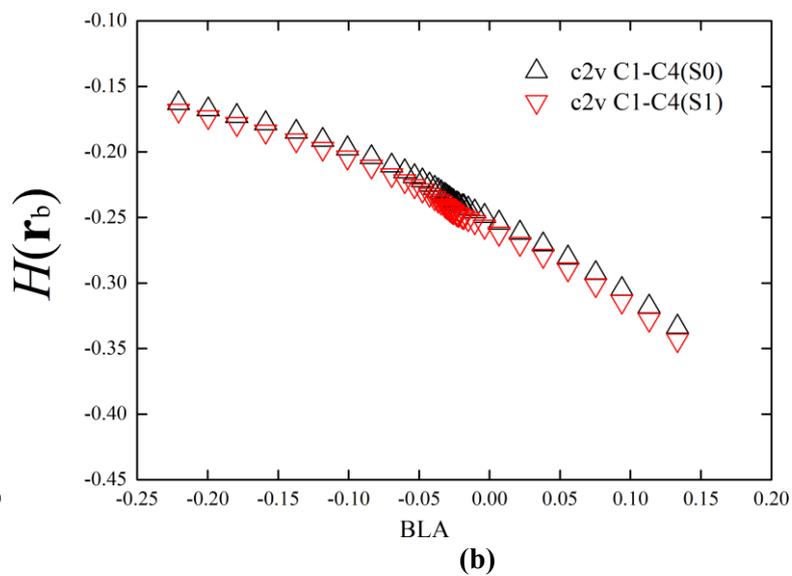

(b)

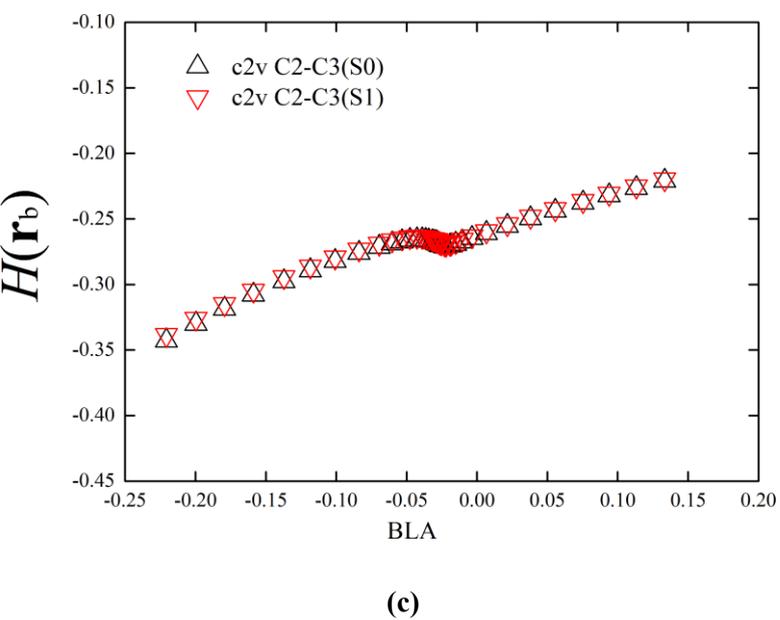

(c)

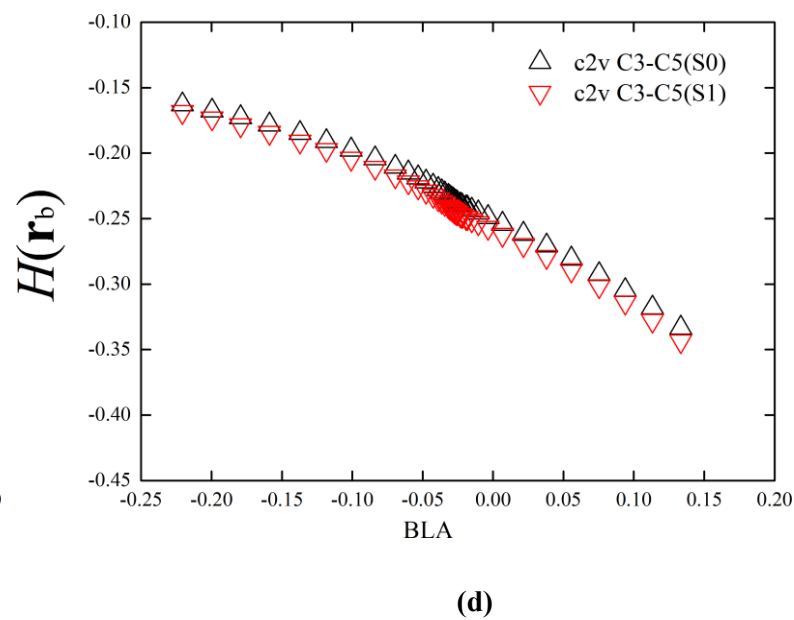

(d)

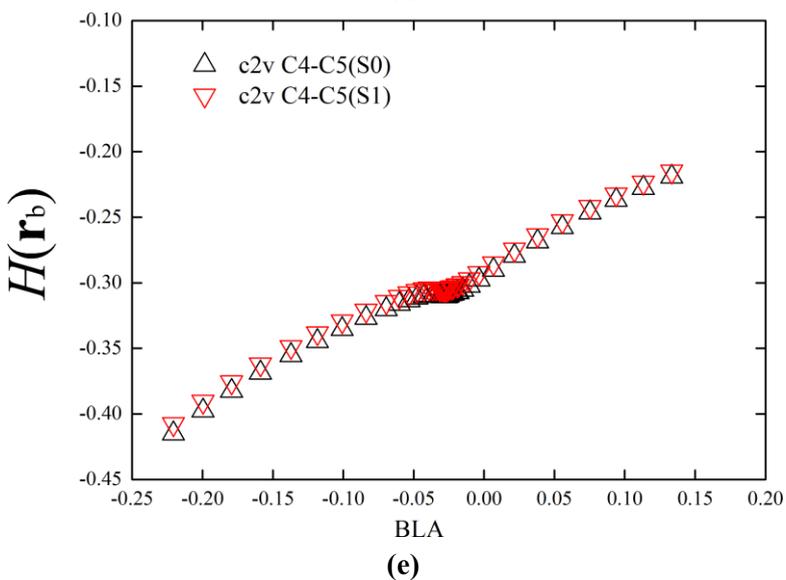

(e)

**Figure S2.** The variation of the $H(\mathbf{r}_b)$ with the BLA for C-C *BCPs* is shown in sub-figures **(a)** - **(e)**.

## 3. Supplementary Materials S3.

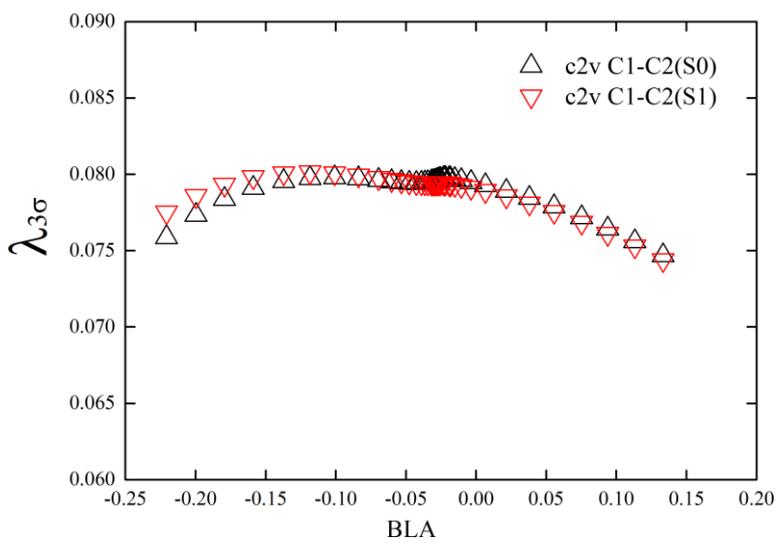

(a)

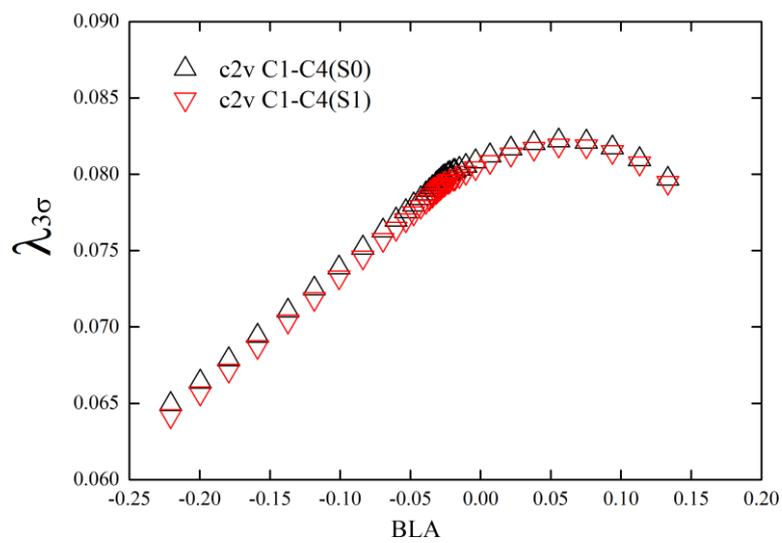

(b)

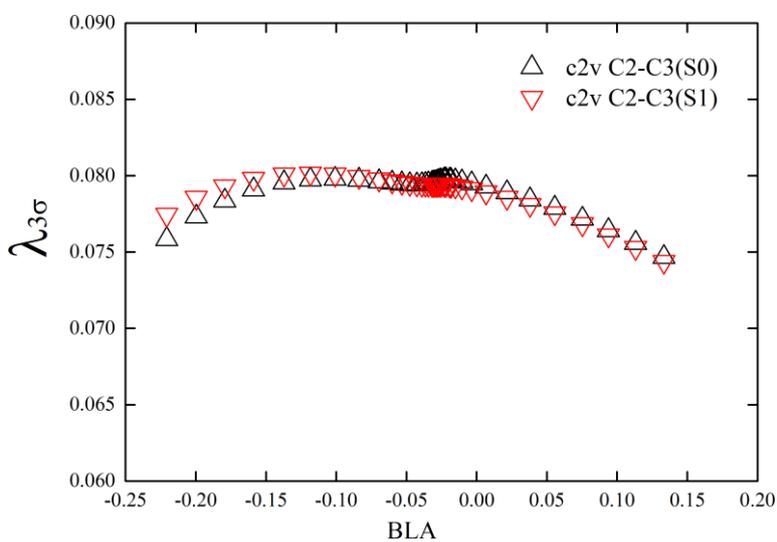

(c)

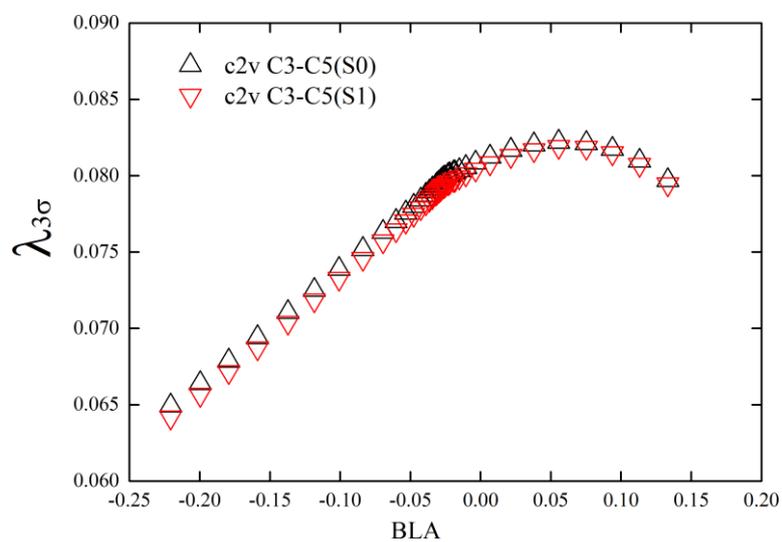

(d)

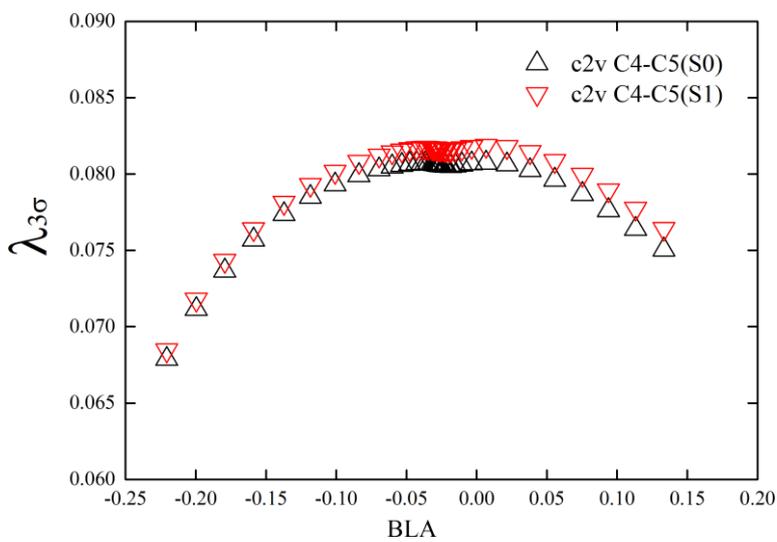

(e)

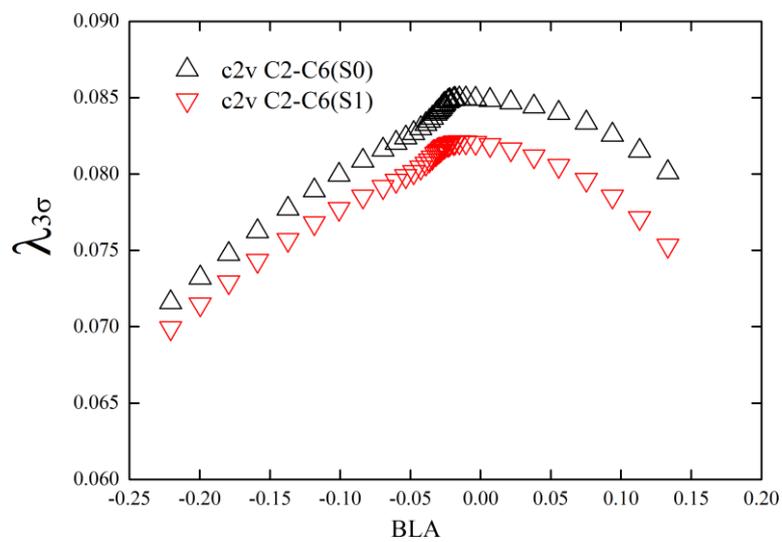

(f)

**Figure S3.** The variation of the stress tensor eigenvalue $\lambda_{3\sigma}$ with the BLA for C-C *BCPs* is shown in sub-figures **(a)** - **(f)**.

## 4. Supplementary Materials S4.

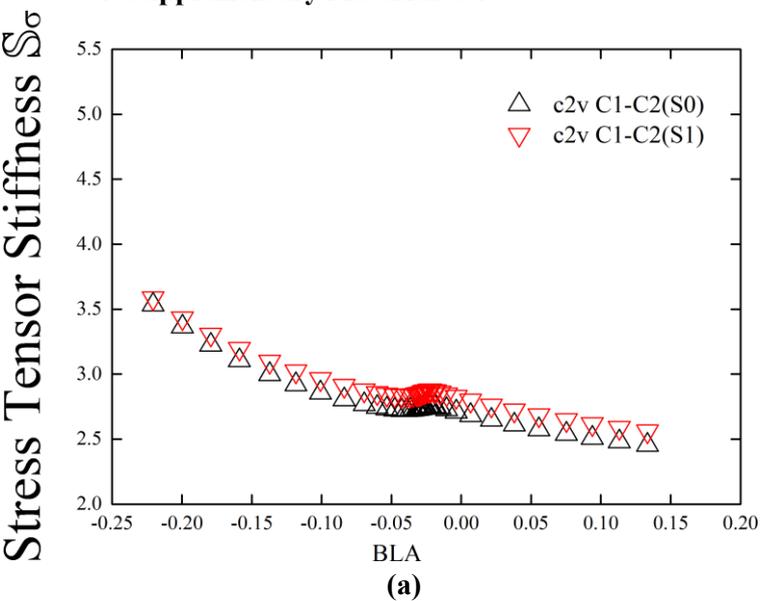

**(a)**

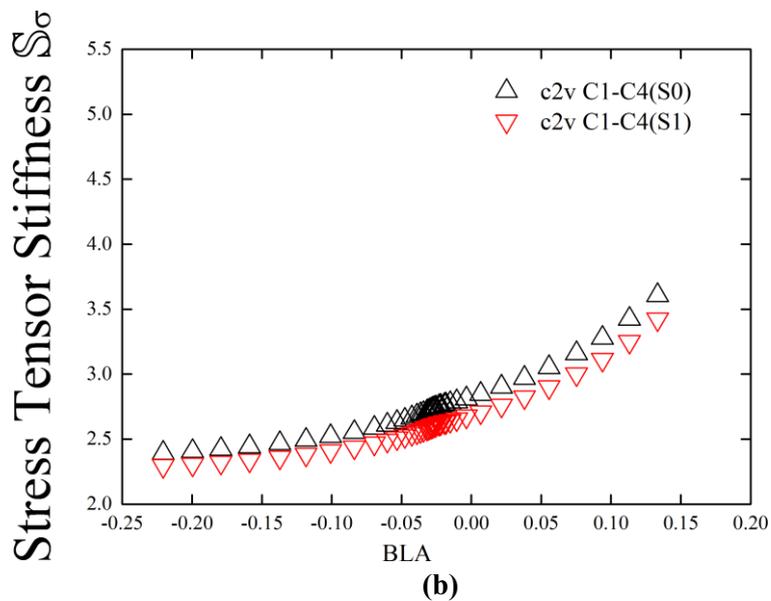

**(b)**

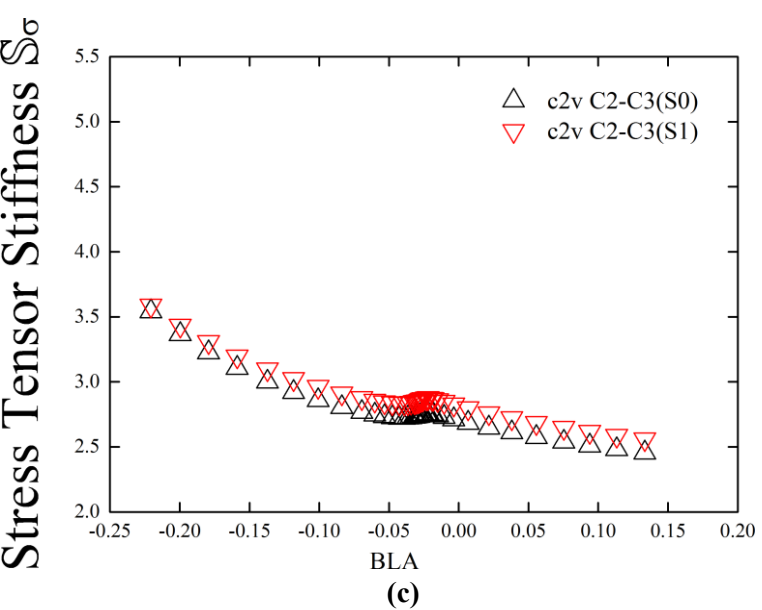

**(c)**

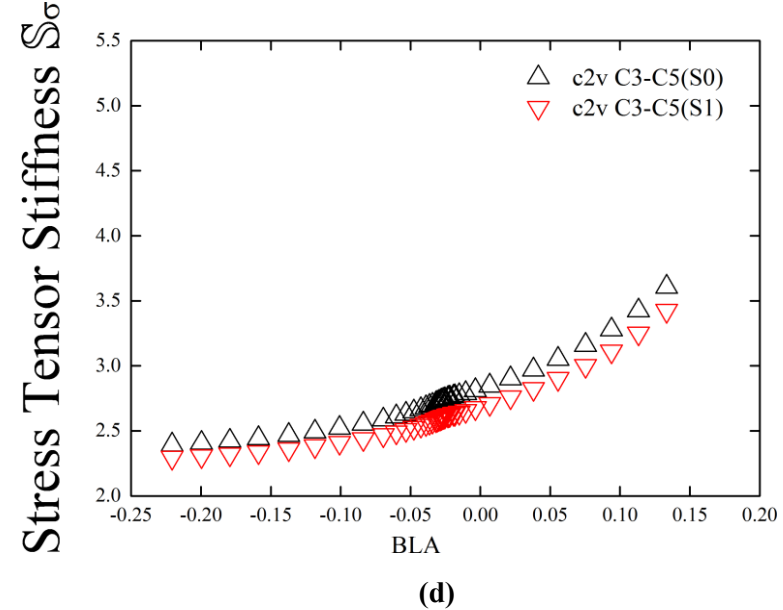

**(d)**

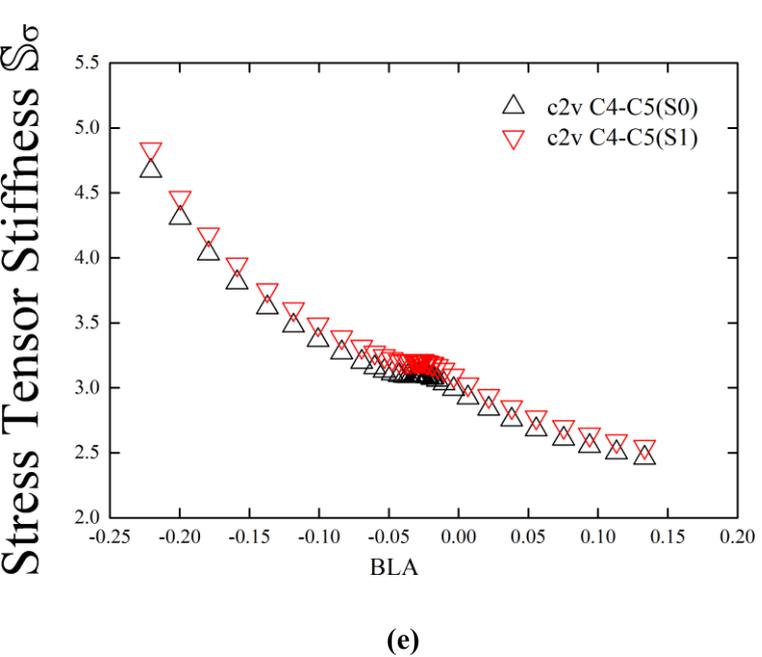

**(e)**

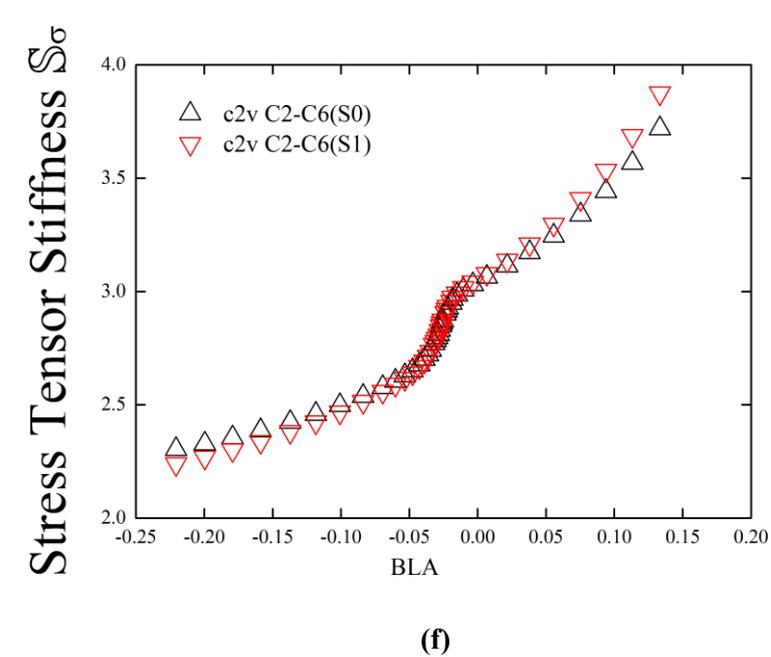

**(f)**

**Figure S4.** The variation of the stress tensor stiffness $\mathbb{S}_\sigma$ with the BLA for C-C *BCPs* is shown in sub-figures **(a)** - **(f)**.

## 5. Supplementary Materials S5.

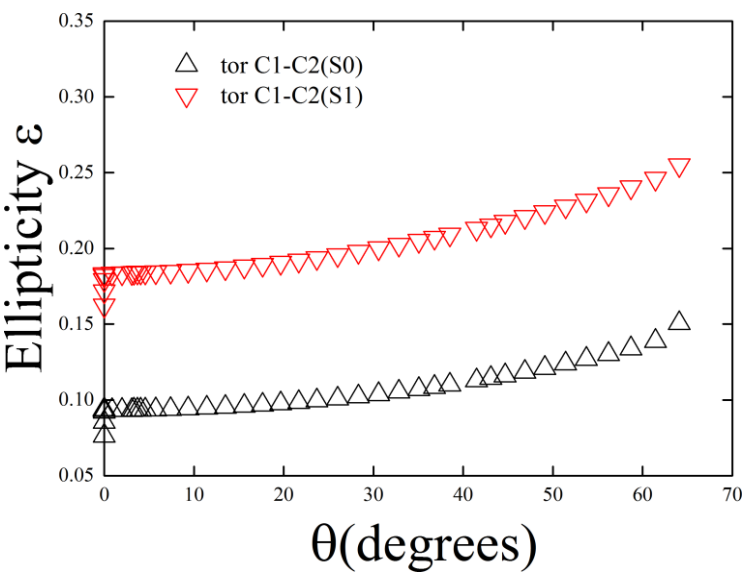

(a)

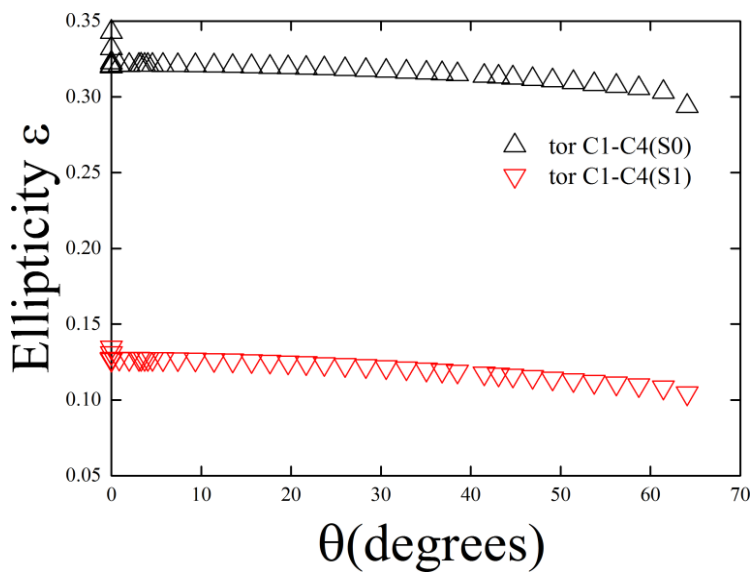

(b)

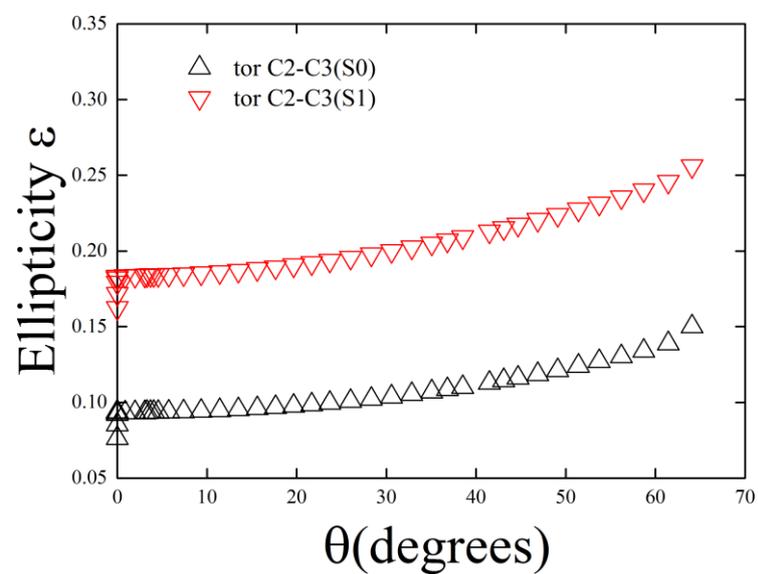

(c)

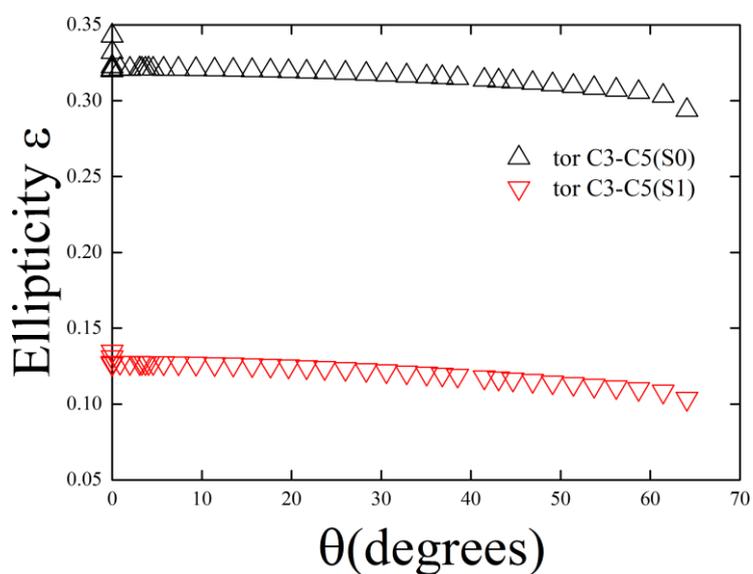

(d)

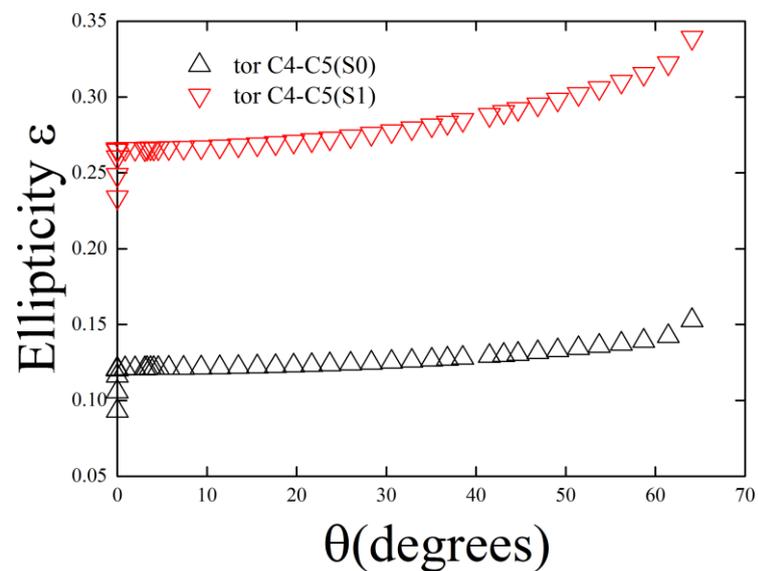

(e)

**Figure S5.** The variation of the ellipticity $\varepsilon$ with the torsion angle $\theta$ for C-C *BCPs* is shown in sub-figures **(a)** - **(e)**.

## 6. Supplementary Materials S6.

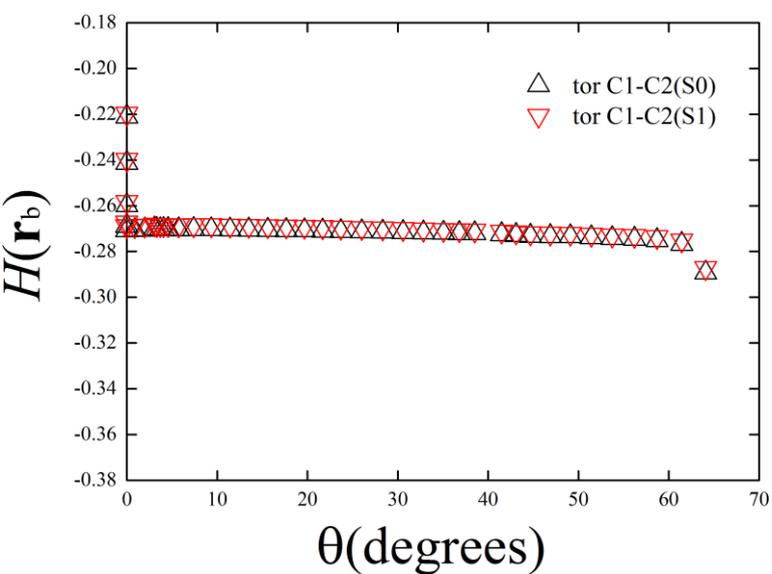

(a)

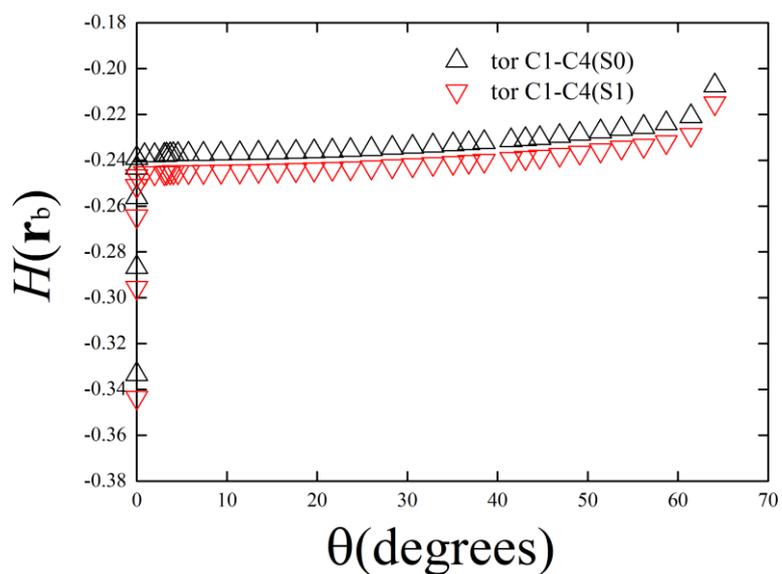

(b)

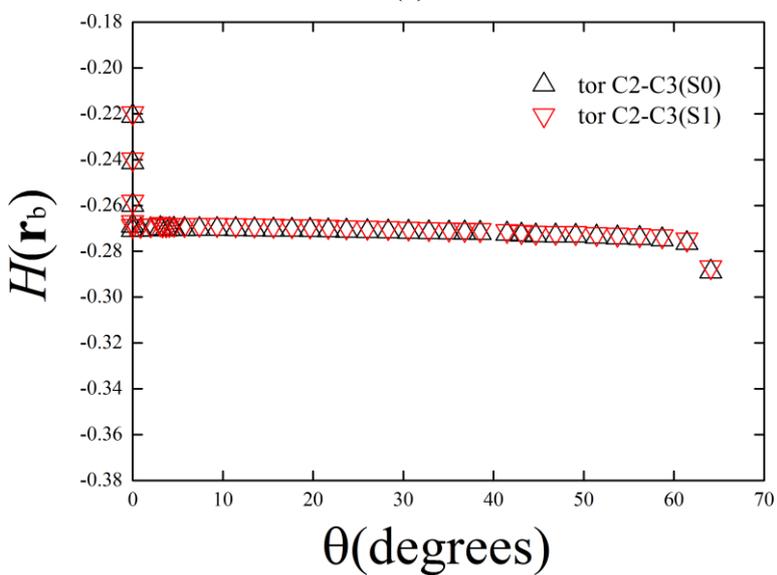

(c)

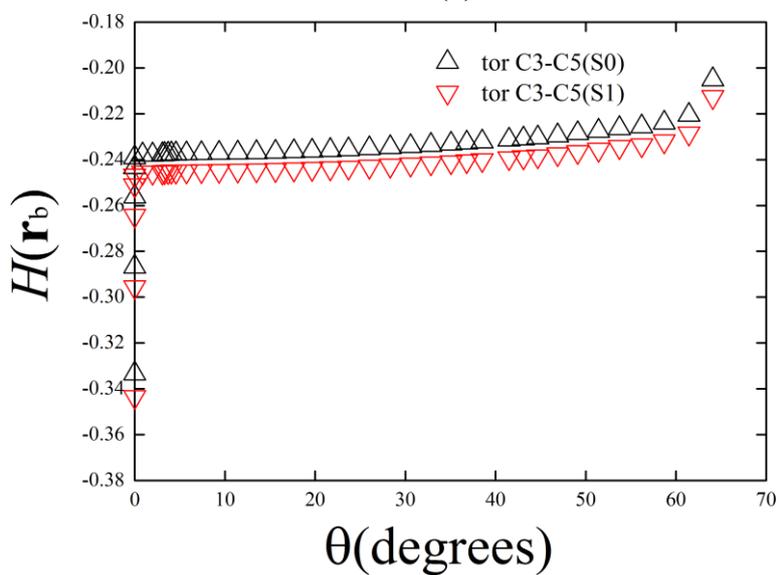

(d)

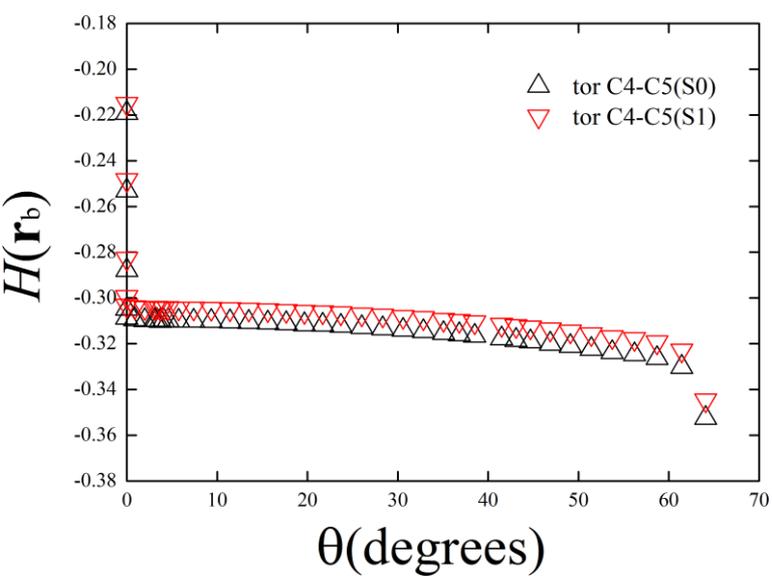

(e)

**Figure S6.** The variation of the $H(\mathbf{r}_b)$ with the torsion angle $\theta$ for C-C *BCPs* is shown in sub-figures **(a)** - **(e)**.

## 7. Supplementary Materials S7.

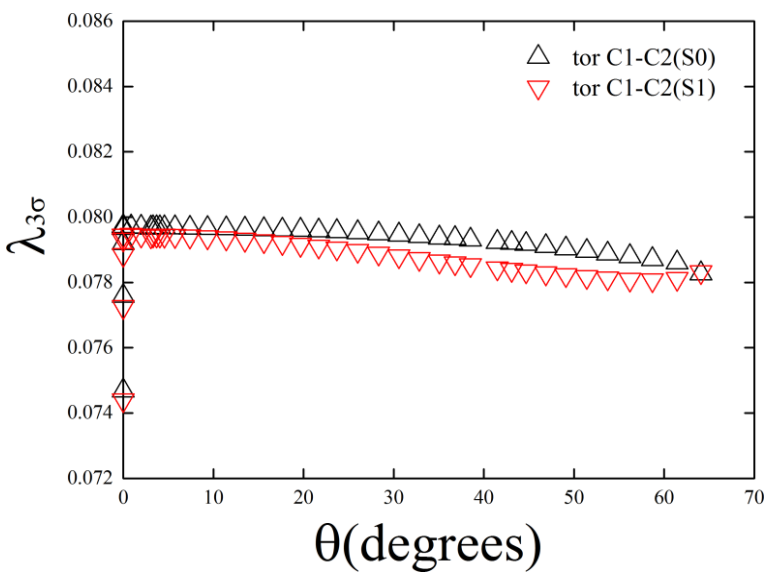

(a)

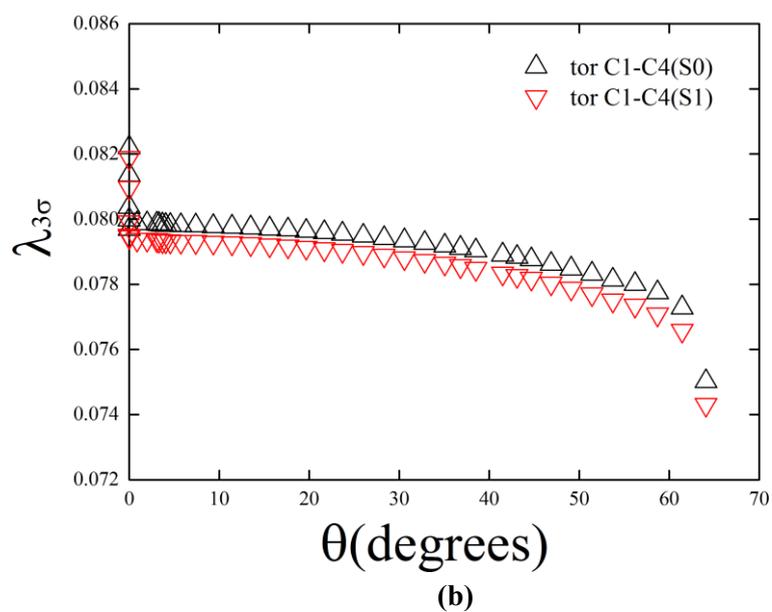

(b)

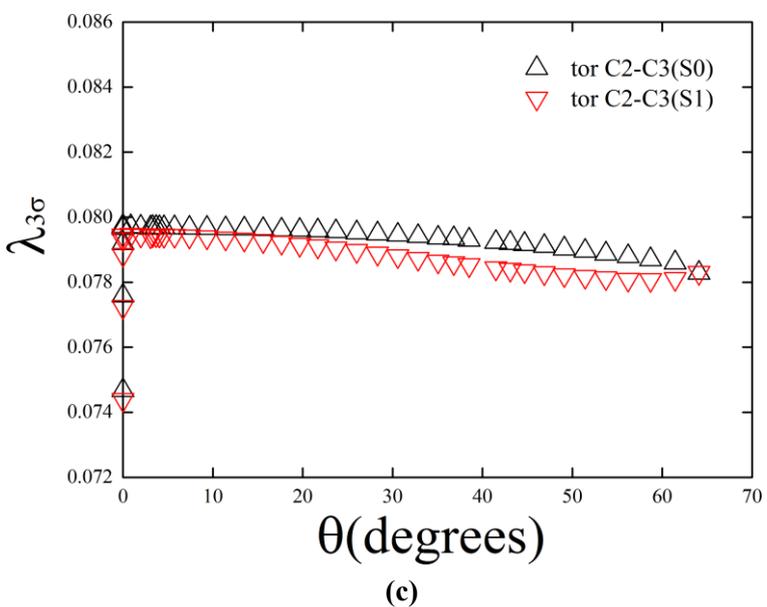

(c)

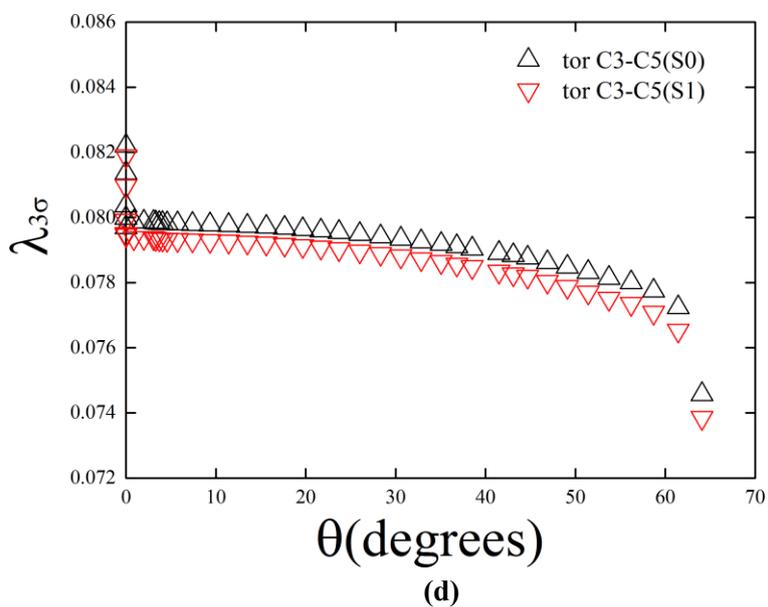

(d)

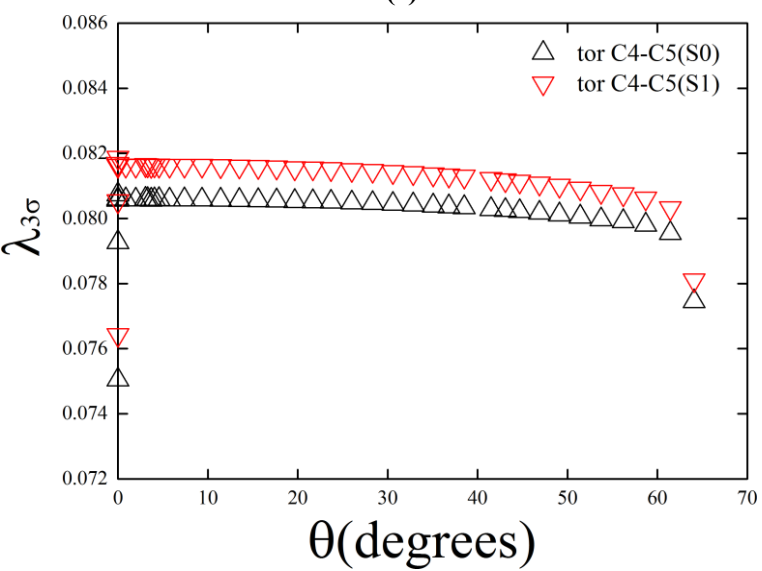

(e)

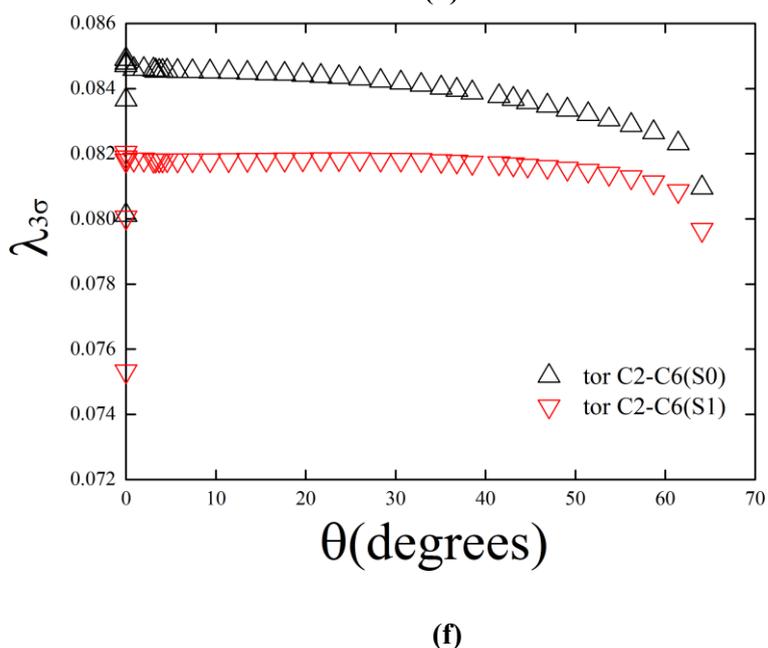

(f)

**Figure S7.** The variation of the stress tensor eigenvalue $\lambda_{3\sigma}$ with the torsion angle $\theta$ for C-C *BCPs* is shown in sub-figures **(a)** - **(f)**.

## 8. Supplementary Materials S8.

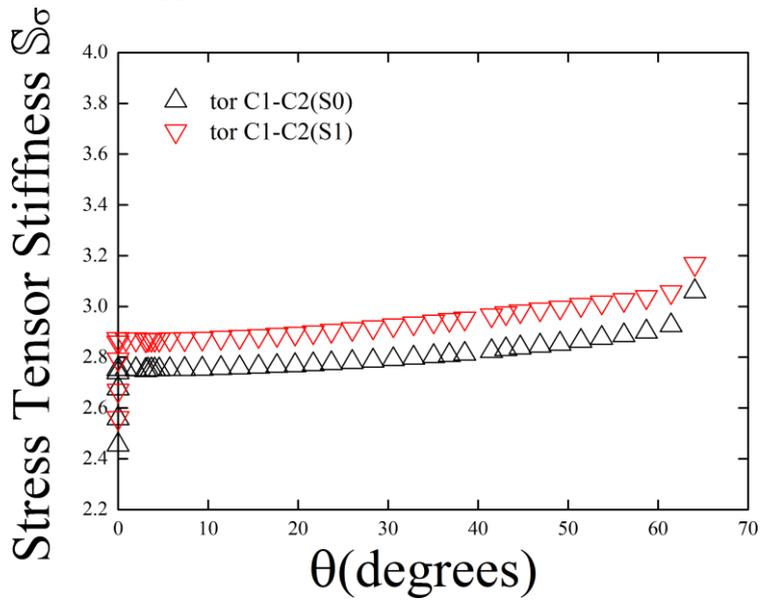

(a)

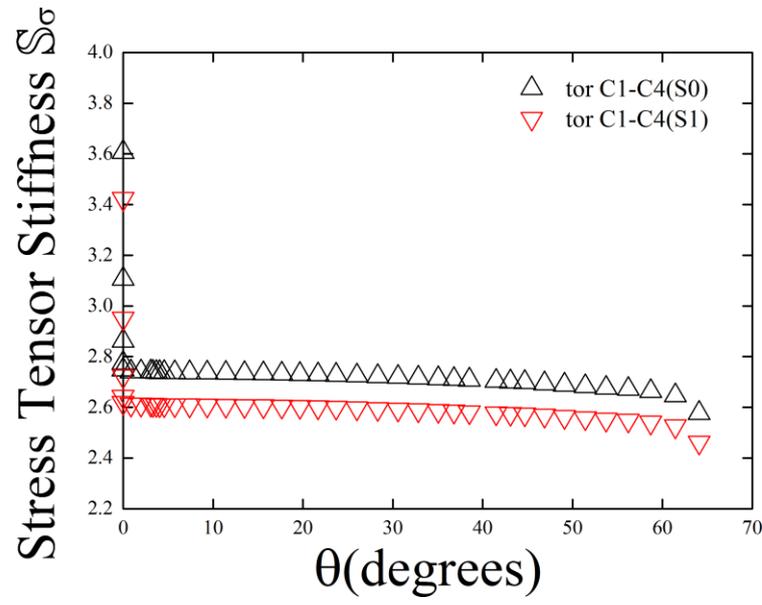

(b)

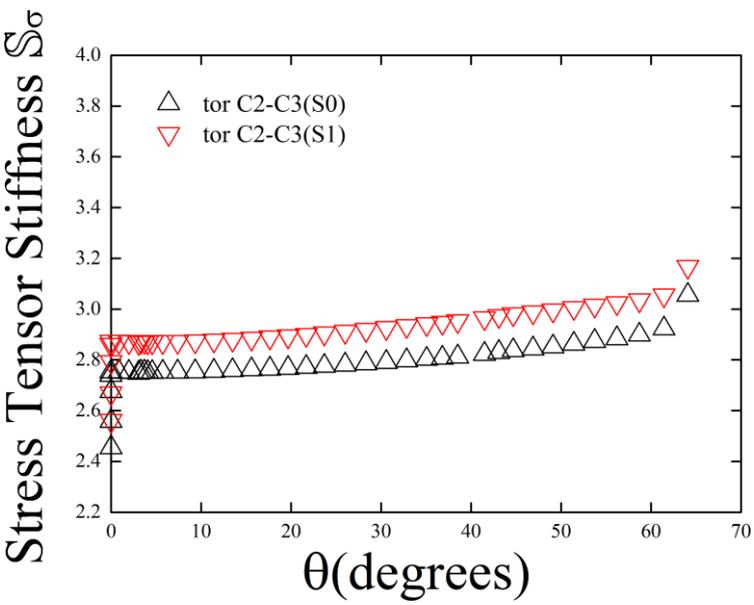

(c)

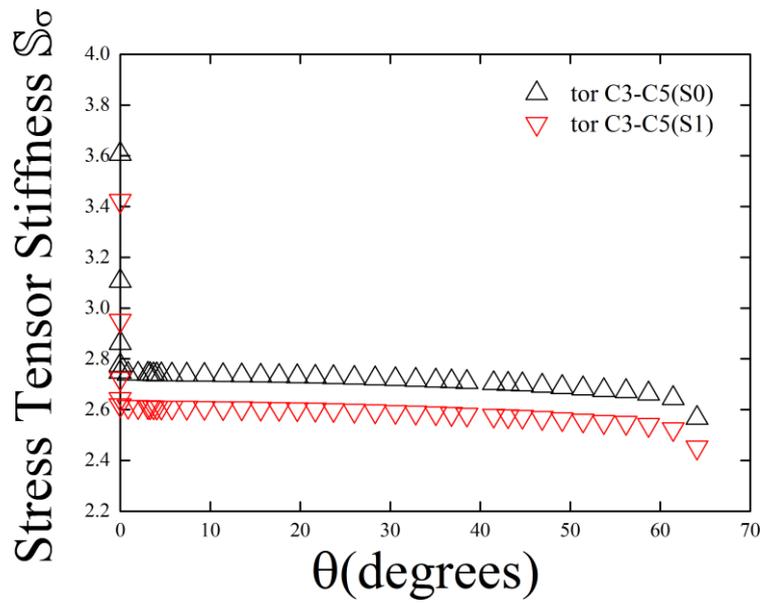

(d)

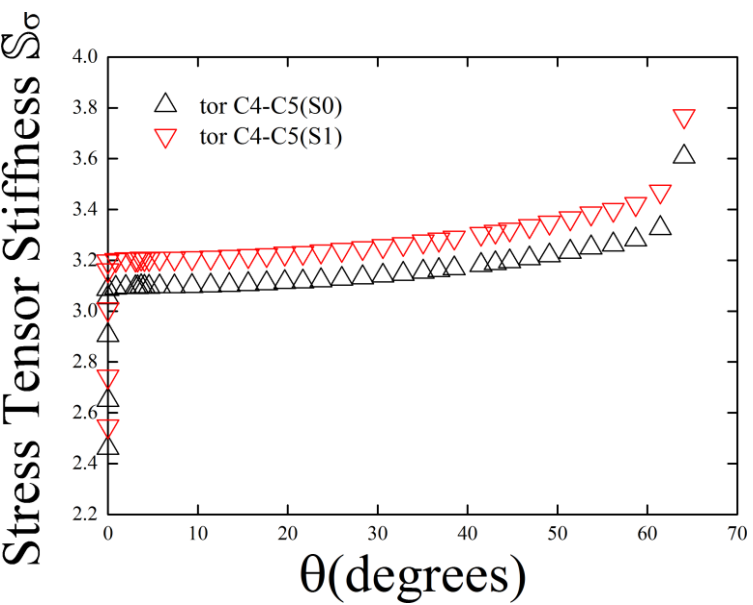 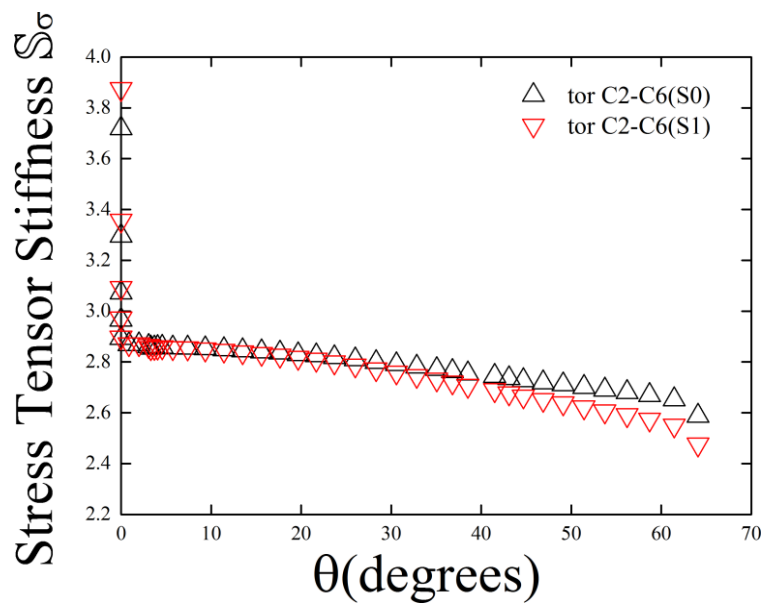

(e) (f)

**Figure S8.** The variation of the stress tensor stiffness $\mathbb{S}_\sigma$ with the torsion angle θ for C-C *BCPs* is shown in sub-figures **(a)** - **(f)**.

## 9. Supplementary Materials S9.

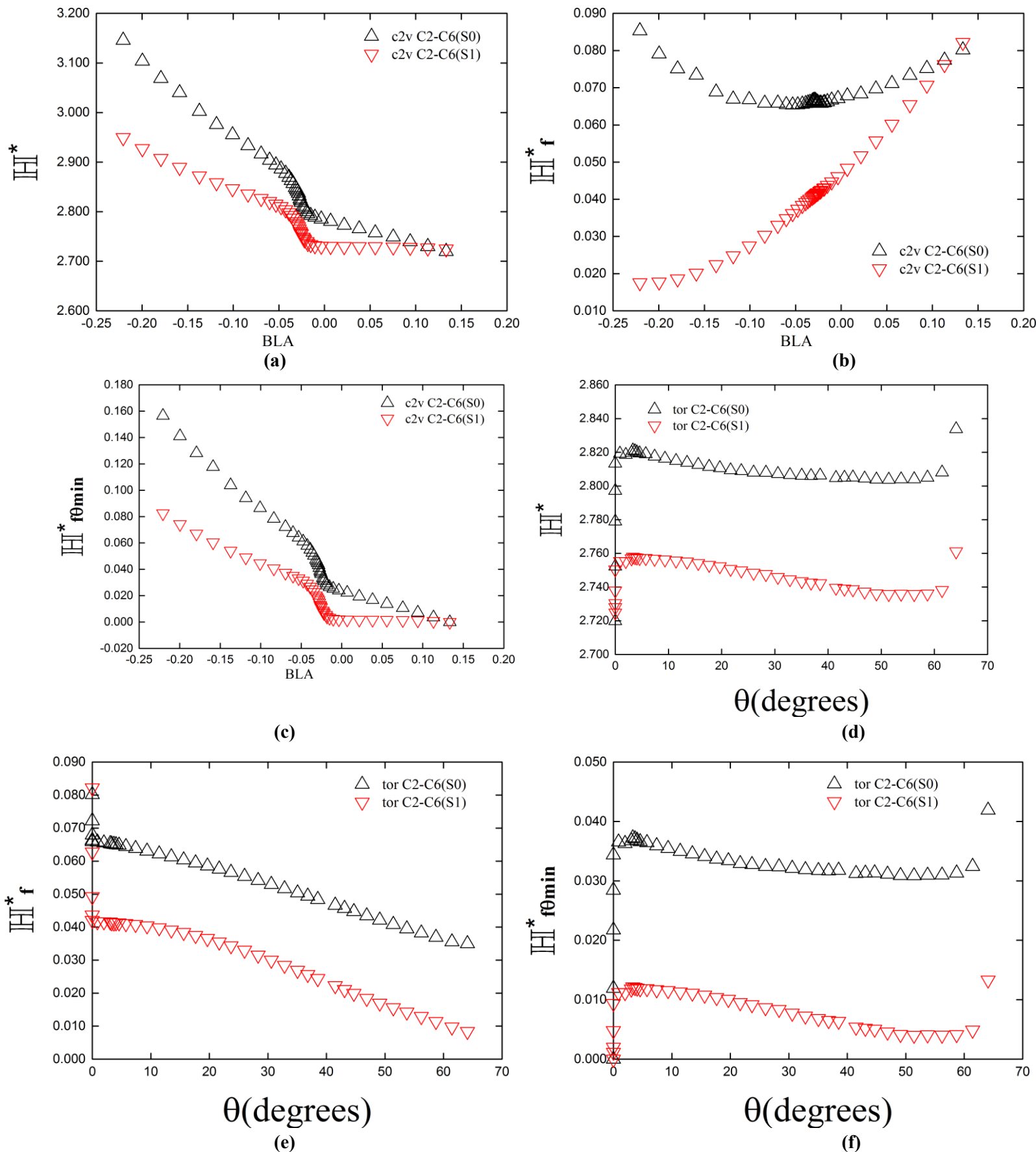

**Figure S9**. The variation of the eigenvector-following path lengths $\mathbb{H}^*$, $\mathbb{H}^*_f = (\mathbb{H}^* - BPL)/BPL$ and $\mathbb{H}^*_{f\theta min} = (\mathbb{H}^* - \mathbb{H}^*_{\theta min})/\mathbb{H}^*_{\theta min}$ of the C2-C6 *BCP* with the BLA are shown in sub-figures **(a)** - **(c)** respectively. The variation of the eigenvector-following path lengths $\mathbb{H}^*$, $\mathbb{H}^*_f = (\mathbb{H}^* - BPL)/BPL$ and $\mathbb{H}^*_{f\theta min} = (\mathbb{H}^* - \mathbb{H}^*_{\theta min})/\mathbb{H}^*_{\theta min}$ of the C2-C6 *BCP* with the torsion θ are presented in sub-figures **(d)** - **(f)** respectively.

## 10. Supplementary Materials S10.
Implementation details of the calculation of the eigenvector-following path lengths $\mathbb{H}$ and $\mathbb{H}^*$.

When the QTAIM eigenvectors of the Hessian of the charge density $\rho(\mathbf{r})$ are evaluated at points along the bond-path, this is done by requesting them via a spawned process which runs the selected underlying QTAIM code, which then passes the results back to the analysis code. For some datasets, it occurs that, as this evaluation considers one point after another in sequence along the bond-path, the returned calculated $\underline{e}_2$ (correspondingly $\underline{e}_1$ is used to obtain $\mathbb{H}^*$) eigenvectors can experience a 180-degree 'flip' at the 'current' bond-path point compared with those evaluated at both the 'previous' and 'next' bond-path points in the sequence. These 'flipped' $\underline{e}_2$ (or $\underline{e}_1$) eigenvectors, caused by the underlying details of the numerical implementation in the code that computed them, are perfectly valid, as these are defined to within a scale factor of -1 (i.e. inversion). The analysis code used in this work detects and re-inverts such temporary 'flips' in the $\underline{e}_2$ (or $\underline{e}_1$) eigenvectors to maintain consistency with the calculated $\underline{e}_2$ (or $\underline{e}_1$) eigenvectors at neighboring bond-path points, in the evaluation of eigenvector-following path lengths $\mathbb{H}$ and $\mathbb{H}^*$.

.